\documentclass[fleqn,usenatbib, useAMS]{mnras}
\usepackage{newtxtext,newtxmath}

\usepackage[T1]{fontenc}
\usepackage{ae,aecompl}

\usepackage{graphicx}	
\usepackage{amsmath}	
\usepackage{amssymb}	
\usepackage{bm}		

\newcommand{\Tr}{\text{Tr}}
\newcommand{\comovhubble}{\mathcal{H}}
\newcommand{\likelihoodfunction}{\mathcal{L}}

\title{Anatomy of Cosmic Tidal Reconstruction}

\author[N. G. Karacayli and N. Padmanabhan]{Naim Goksel Karacayli$^{1}$ and Nikhil Padmanabhan$^{1, 2}$
\\
$^{1}$Department of Physics, Yale University, New Haven, CT, USA\\
$^{2}$Department of Astronomy, Yale University, New Haven, CT, USA}

\date{\today}

\pubyear{2018}

\begin{document}
\label{firstpage}
\pagerange{\pageref{firstpage}--\pageref{lastpage}}
\maketitle

\begin{abstract}
21-cm intensity surveys aim to map neutral hydrogen atoms in the universe through hyper-fine emission. Unfortunately, long-wavelength (low-wavenumber) radial modes are highly contaminated by smooth astrophysical foregrounds that are six orders of magnitude brighter than the cosmological signal. This contamination also leaks into higher radial and angular wavenumber modes and forms a foreground wedge. Cosmic tidal reconstruction aims to extract the large-scale signal from anisotropic features in the local small-scale power spectrum through non-linear tidal interactions; losing small-scale modes to foreground wedge will impair its performance. In this paper, we review tidal interaction theory and estimator construction, and derive the theoretical expressions for the reconstructed spectra. We show the reconstruction is robust against peculiar velocities. Removing low line-of-sight $k$ modes, we demonstrate cross-correlation coefficient $r$ is greater than 0.7 on large scales ($k \la 0.1$ $h/$Mpc) even with a cutoff value $k^c_{\|}=0.1$ $h/$Mpc. Discarding wedge modes yields $0.3\la r \la 0.5$ and completely removes the dependency on $k^c_{\|}$. Our theoretical predictions agree with these numerical simulations. 
\end{abstract}




\section{Introduction}
\label{sec:intro}
21-cm intensity mapping has been a promising probe for cosmological information on large-scales \citep{bull_late-time_2015}. Late-time 21-cm measurements such as CHIME \citep{bandura_canadian_2014} and HIRAX \citep{newburgh_hirax:_2016} aim to map large volumes of the universe through red-shifted neutral hydrogen emission line in $0.8 < z < 2.5$ redshift range. This method faces challenges from foregrounds that are six orders of magnitude brighter than the cosmological signal. Long-wavelength modes in the line-of-sight are contaminated by these spectrally smooth foregrounds. Furthermore, instrumental imperfections cause this noise to leak into high $k_{\bot} - k_{\|}$ modes and to form a foreground wedge in Fourier space \citep{morales_four_2012}. Since the long-wavelength radial modes play a significant role in cross correlations of 21-cm surveys with CMB measurements and photo-z galaxy surveys \citep{furlanetto_cross-correlation_2007, adshead_reionization_2008, masui_measurement_2013}, recovering the lost modes is important to improve the cross correlation signal. 

A procedure called cosmic tidal reconstruction has been developed to extract these lost modes from small-scale signal \citep{pen_cosmic_2012, zhu_cosmic_2016, zhu_recovering_2018}. Although large-scale modes are not explicitly present, they leave an anisotropic imprint on small-scale modes through non-linear tidal interactions. To better visualize the nature of this interaction, we remind the reader the Moon's tidal field on Earth. The gravitational potential of the Moon around the Earth's centre can be Taylor expanded up to second order. The constant term can be ignored; the first derivative is the net gravitational force between two celestial objects. The second derivative of the potential causes deformations on Earth's oceans and is the tidal field. Similarly, the cosmological gravitational potential can be separated into short- and long-wavelength parts. The latter can be expanded up to second order around an observer. This tidal field $t_{ij} \sim \Phi_{L, ij}$ then causes deformations in the local universe and introduces anisotropies to the local small-scale correlation function and power spectrum.

After finding the theoretical expression for the local small-scale power spectrum, we now need to estimate the underlying large-scale over-density field $\delta_L$ given an observation $\delta$. Assigning a Gaussian probability to measuring a data set $\delta$ for a given $\delta_L$, we can claim that our best estimate maximizes this probability. In technical terms, optimal quadratic estimators for $\delta_L$ are constructed assuming the observed density field $\delta$ is Gaussian. This assumption mandates mapping $\delta$ to a Gaussian field. 
Choosing robust $t_{ij}$ components is the last piece of the puzzle. Galaxies have peculiar velocities on top of the Hubble expansion. The line-of-sight positions are mismeasured because of these peculiar velocities. To minimize the so called redshift space distortions, two quadrupolar distortions of the tidal field in the plane perpendicular to the line-of-sight ($\gamma_1 = (\Phi_{L, xx} - \Phi_{L, yy})$ and $\gamma_2 = 2\Phi_{L, xy}$) are chosen. Even though both $\gamma_1$ and $\gamma_2$ are independent estimators of $\delta_L$ in theory, they can be combined into the three dimensional convergence field $\kappa_{3D}$ which is a better estimate for $\delta_L$ given that each $\gamma$ is zero in distinct $\hat{k}$ direction \citep{kaiser_weak_1992}. A Wiener filter finally corrects for the bias and noise in $\kappa_{3D}$ itself. 

A reconstruction algorithm first has to pass a test on ideal input by producing a highly correlated field. Then, it has to be robust against the obstacles arising from imperfections. \citet{zhu_cosmic_2016} showed cosmic tidal reconstruction produced cross-correlations greater than 0.9 on scales $k \la 0.1$ $h/$Mpc at $z=0$ for full dark matter field in real space. In this paper, we run the cosmic tidal reconstruction in real and redshift spaces to test the efficiency of the algorithm. We introduce the observational challenges to assess if the reconstruction is robust. In short, our work includes redshift space distortions, testing a range for low $k_{\|}$ data loss and including the foreground wedge. 

The reader may be wondering how all these steps produce a correlated signal. The theoretical expectations are concealed behind tidal interactions, estimators and multiple Fourier transforms in implementation. Because of these complications, previous works relied only on simulations. In this work, we derive the analytic expressions for cross and power spectrum. We show that the cross correlations emerge from a modified bispectrum. This modification is due to Gaussian mapping of the initial over-density field. Our theoretical model can quantify the steep decline in reconstruction efficiency due to lost modes and the minor degradation due to peculiar velocities. It also promises a Wiener filter estimate when higher order terms are considered. 

In Section~\ref{sec:theory}, we review the tidal interaction theory and the construction of estimators. Section~\ref{sec:implementation} describes the parameters for our N-body simulations and outlines the algorithm step by step. We present our results from these simulations in Section~\ref{sec:results}. We derive analytic cross and auto spectrum expressions in Section~\ref{sec:reconst_theory}; and summarize in Section~\ref{sec:summary}.



\section{Theory}
\label{sec:theory}
In this section, we first review the derivation of the local small-scale power spectrum  \citep{zhu_cosmic_2016, schmidt_large-scale_2014} and connect it to the construction of estimators \citep{lu_precision_2008}. We refer the reader to the references for a longer discussion. Moreover, \citet{akitsu_large-scale_2017} and  \citet{akitsu_impact_2018} study the impact of the large-scale tidal field in a similar but different framework.

\subsection{Local Power Spectrum}
The first ingredient of the cosmic tidal reconstruction is a theoretical prediction for local anisotropic features. What we call tidal interactions are the mechanism introducing these anisotropies. \citeauthor{zhu_cosmic_2016} has reviewed the traceless tidal field, and \citeauthor{schmidt_large-scale_2014} has studied the tidal interaction theory in detail using conformal Fermi Normal Coordinate frame ($\overline{\text{FNC}}$). We find it sufficient to work in Newtonian picture for our discussion. Suppressing the time dependence, we start by decomposing the gravitational potential in Fourier space to separate it into short- and long-wavelength parts.
\begin{align}
	\phi(\textbf{x}) &= \int \frac{d^3 \textbf{k}}{(2\pi)^3} \phi(\textbf{k}) e^{i\textbf{k}\cdot \textbf{x}} \\
    &= \int\displaylimits_{k<k_L} \frac{d^3 \textbf{k}}{(2\pi)^3} \phi(\textbf{k}) e^{i\textbf{k}\cdot \textbf{x}} + \int\displaylimits_{k>k_L} \frac{d^3 \textbf{k}}{(2\pi)^3} \phi(\textbf{k}) e^{i\textbf{k}\cdot \textbf{x}} \\
    &= \Phi_L(\textbf{x}, k_L) + \Phi_S(\textbf{x}, k_L)
\end{align}

Now we consider a spherical volume of radius R around the origin such that $k_L R \la 1$. Suppressing $k_L$ we can rewrite the long-wavelength potential as
\begin{align}
	\Phi_L(\textbf{x}) &= \int\displaylimits_{k<k_L} \frac{d^3 \textbf{k}}{(2\pi)^3} \phi(\textbf{k})  e^{i\textbf{k}\cdot \textbf{x}} \\
    &= \int\displaylimits_{k<k_L} \frac{d^3 \textbf{k}}{(2\pi)^3} \phi(\textbf{k})  \left(1 + i\textbf{k}\cdot \textbf{x} -\frac{1}{2}k_i k_j x^i x^j \right) \\
    &= \Phi_L(\textbf{0}) + \textbf{x} \cdot \nabla \Phi_L(\textbf{0}) + \frac{1}{2}x^i x^j \Phi_{L, ij} (\textbf{0}).
\end{align}
We can drop the constant term in the potential. The first derivative is a net force on the local universe; for large enough scales this can be ignored as well. The trace of $\Phi_{L, ij} (\textbf{0})$ describes if the local universe lives at an under- or over-dense region. The effect of this term is an overall change to the growth rate, which is not detectable and we ignore its contribution here. So with a traceless tidal field $t_{ij}(\tau) \equiv \Phi_{L, ij}(\textbf{0}) - \delta^K_{ij}\Phi_{L, kk}(\textbf{0})/3$, the gravitational potential $\phi$ is given by
\begin{equation}
	\phi(\textbf{x}) = \Phi_S + \epsilon_t\frac{1}{2}t_{ij}(\tau)x^ix^j,
\end{equation}
where small-scale gravitational potential $\Phi_S$ obeys the Poisson equation $\nabla^{2}_{x} \Phi_S =\frac{3}{2} \Omega_{m}(\tau) \mathcal{H}^{2}(\tau) \delta_S$. We will suppress subscript $S$ to simplify our notation. We added $\epsilon_t$ to keep track of the long-wavelength perturbations in powers of $t_{ij}$. The tidal field can be written as $t_{ij}(\tau) = T(\tau) t^{(0)}_{ij}$, where $t^{(0)}_{ij}$ is the present value, $T(\tau) = D(\tau)/a(\tau)$ and $D(\tau)$ is the linear growth function.

The equation of motion for a particle in an expanding universe is
\begin{equation}
	\left[\frac{d^2}{d\tau^2} + \comovhubble \frac{d}{d\tau}\right]\textbf{x} \equiv \mathbb{D} \textbf{x}= -\nabla_x \phi,
\end{equation}
where we have defined the operator $\mathbb{D}$. We can solve this equation using Lagrangian perturbation theory up to the nearest order in non-linear coupling between long- and short- wavelength perturbations.
\begin{align}
	\textbf{x} &= \textbf{q} + \Psi(\textbf{q}, \tau)\\
    \Psi &= \epsilon_{s}\Psi^{1s} + \epsilon_{t}\Psi^{1t} + \epsilon_{s}\epsilon_{t}\Psi^{1st} + \cdots 
\end{align}
The hypothetical particle in question still lives in the spherical volume, so $x<R$ by construction. Additionally, $1s$ term is the linear solution and $\epsilon_t$ signifies the contribution to the local small-scale fluctuations from the tidal field. We introduced $\epsilon_s$ to keep track of the coupling order.

Let us first expand the over-density field using the mass conservation relation.
\begin{align}
    \delta &= \frac{1}{\det\left( \delta^K_{ij} + \Psi_{i, j} \right)} - 1 \\
    &= -\Tr \Psi_{i, j} + \frac{1}{2}[(\Tr \Psi_{i, j})^2 + \Tr (\Psi_{i, j}^2)],
\end{align}
where $\partial^q_j \Psi_i \equiv \Psi_{i, j}$. We set $\Psi^{1t}_{i, i}=0$ with hindsight because the tidal field is traceless ($t_{ii}=0$). The corresponding over-density in Lagrangian coordinates is
\begin{equation}
	\delta_{\text{LG}}(\textbf{q}) = \epsilon_{s}\delta_{1s} + \epsilon_{s}\epsilon_{t}\left[\Psi_{i,j}^{1s}\Psi^{1t}_{j,i} - \Psi^{1st}_{i, i} \right].
\end{equation} 
We find the over-density field in Eulerian frame $\delta_{\text{EU}}$ by demanding $\delta_{\text{LG}}(\textbf{q}) = \delta_{\text{EU}}(\textbf{x})$.
\begin{align}
	\delta_{\text{EU}}(\textbf{x}) &= \delta_{\text{LG}}(\textbf{x} - \Psi) \\
    &= \delta_{\text{LG}}(\textbf{x}) - \Psi \cdot \nabla \delta_{\text{LG}} (\textbf{x}) \\
    \delta_{\text{EU}}(\textbf{x}) &= \epsilon_{s}\delta_{1s}(\textbf{x}) + \epsilon_{s}\epsilon_{t} \left[\Psi_{i,j}^{1s}\Psi^{1t}_{j,i} - \Psi^{1st}_{i, i} \right]({\textbf{x}}) - \epsilon_{s}\epsilon_{t}\Psi_i^{1t} \partial_i \delta_{1s}(\textbf{x})
\end{align}
Then, the tidal field's contribution to the local density fluctuations $\delta_t = \delta_{\text{EU}} - \delta_{1s}$ is
\begin{align}
	\delta_t(\textbf{x}) &= \left[\Psi_{i,j}^{1s}\Psi^{1t}_{j,i} - \Psi^{1st}_{i, i} \right]({\textbf{x}}) - \Psi_i^{1t} \partial_i \delta_{1s}({\textbf{x}}).
\end{align}

Now, we go back to the equation of motion. We find the divergence using the chain rule $\partial^q_i = \partial^x_i + \Psi_{j,i}\partial^x_j$. 
\begin{align}
    \mathbb{D}\Psi_{i,i} &= -\nabla_x^2 \Phi - \Psi_{j,i}(\partial^x_j \partial^x_i \Phi + \epsilon_t t_{ij})
\end{align}
We can solve for the displacement field order by order. Here, we only summarize the results.
\begin{align}
    \Psi^{1s}_i &= - D(\tau) \frac{\partial^q_i}{\nabla^2_q} \delta^{(0)}_{1s} \\
    \Psi^{1t}_i &= - F(\tau)t^{(0)}_{ij}q_j \\
    \Psi^{1st}_{i,i} &= D_{1st}(\tau) D(\tau) t^{(0)}_{ij} \frac{\partial^q_i \partial^q_j}{\nabla^2_q} \delta^{(0)}_{1s},
\end{align}
and the time evolution functions are given by
\begin{align}
    F(\tau) &= \int_0^{\tau} d\tau^{\prime\prime} D(\tau^{\prime\prime}) \int_{\tau^{\prime\prime}}^{\tau} \frac{d\tau^{\prime}}{a(\tau^{\prime})} \\
    D_{1st}(\tau) &= \int_0^{\tau}\!\!\! d\tau^{\prime} \frac{H(\tau)D(\tau^{\prime}) - D(\tau) H(\tau^{\prime})}{\dot H(\tau^{\prime}) D(\tau^{\prime}) - H(\tau^{\prime}) \dot D(\tau^{\prime})} \frac{T(\tau^{\prime})D(\tau^{\prime})}{D(\tau)}.
\end{align}
These functions in terms of $z$ integral are in Appendix~\ref{app:time_dep_fn}.

Putting all these together and defining $\alpha(\tau) \equiv -D_{1st}(\tau) + F(\tau)$, we find $\delta_t$ in terms of $\delta_{1s}$ and its spatial derivatives.
\begin{equation}
    \delta_t = t^{(0)}_{ij}\left[ \alpha(\tau) \frac{\partial_i\partial_j}{\nabla^2} + F(\tau)x_j\partial_i\right]\delta_{1s}(\textbf{x},\tau).
\end{equation}

The local small-scale correlation function is defined with respect to a local origin, $\xi(\textbf{r}) = \langle \delta(\textbf{0}) \delta(\textbf{r}) \rangle$ where $\delta = \delta_{1s}+\delta_t$. Then, the nearest order power spectrum  under tidal distortions is
\begin{equation}
	\tilde{P}_{1s}(\textbf{k},\tau) = P_{1s}(k,\tau) + t_{ij}^{(0)} \hat{k}^{i} \hat{k}^{j} f(k,\tau) P_{1s}(k,\tau)\label{eq:p1stilde},
\end{equation}
where $f(k, \tau) = -2\alpha(\tau) - F(\tau) d\ln P_l(k,\tau)/d\ln k$. For clarity, we stress that $1s$ term is the linear solution and $P_{1s}\equiv P_l$ is the linear power spectrum. Tilde represents tidal distortions, and $\hat{k}_i=k_i/k$. 

Analogous to weak lensing, we define $\gamma_1 = (\Phi_{L, xx} - \Phi_{L, yy})$ and $\gamma_2 =2 \Phi_{L, xy}$ and obtain 
\begin{align}
    \Delta P_{1s} &= f(k, z) P_{l}(k,z) \left[ (\hat{k}^{2}_{x} - \hat{k}^{2}_{y}) \gamma_{1}^{(0)} + 2 \hat{k}_{x}\hat{k}_{y}\gamma_{2}^{(0)} \right].
\end{align}
Note that we ignore the contribution from the $z$-components of the tidal tensor to minimize the effects of redshift space distortions on our measurements. 

\subsection{Estimators}
\label{subsec:estimator}
To estimate the underlying long-wavelength over-density field, we assign a Gaussian probability distribution for measuring a data set $\tilde{\delta}(\textbf{k})$ that depends on parameters $\gamma_1$ and $\gamma_2$ \citep{lu_precision_2008}: $P(\tilde{\delta}(\textbf{k}); \gamma_1, \gamma_2) \propto |\mathbb{C}|^{-1/2}\exp(-\tilde{\delta}^{\dagger}\mathbb{C}^{-1}\tilde{\delta}/2)$. However, we pursue the best $\gamma$ estimate of one data set. We define the likelihood function $\likelihoodfunction$ as the negative logarithm of the probability distribution over parameters $\gamma$ for a fixed $\tilde{\delta}(\textbf{k})$. We also assume a diagonal covariance matrix such that $\mathbb{C}_{ij} = \langle \tilde{\delta}(k_i)\tilde{\delta}(k_j) \rangle - \langle \tilde{\delta}(k_i)\rangle \langle\tilde{\delta}(k_j) \rangle = L^3\delta_{ij}\tilde{P}_{tot}(k_i)$. Then, the likelihood function in the continuum limit is
\begin{align}
	\likelihoodfunction &= \int d^{3} k \left[ \ln \tilde{P}_{tot}(\textbf{k}) +\frac{|\tilde{\delta}(\textbf{k})|^{2}}{L^{3} \tilde{P}_{tot}(\textbf{k})} \right],
\end{align}
where an ensemble average gives $\tilde{P}_{tot}(\textbf{k}) = \tilde{P}_{1s}(\textbf{k})  + P_{N}(\textbf{k})$ and equation~(\ref{eq:p1stilde}) provides the theoretical prediction for $\tilde{P}_{1s}(\textbf{k})$. The noise spectrum $P_{N}(\textbf{k})$ includes the effects of unmodelled non-linearities as well as instrumental noise.

We can construct the estimators by maximizing $\likelihoodfunction$ with respect to $\gamma_{1,2}$. Let us go through the calculation for $\gamma_1$.
\begin{align}
	\frac{\partial \likelihoodfunction}{\partial \gamma_1} &= \int \frac{d^{3} k}{(2\pi)^{3}} \left[ \frac{ \tilde{P}_{tot}(\textbf{k}) -|\tilde{\delta}(\textbf{k})|^{2}L^{-3}}{ \tilde{P}^{2}_{tot}(\textbf{k})} \right] \frac{\partial \tilde{P}_{1s}}{\partial \gamma_{1}} \\
    \frac{\partial \tilde{P}_{1s}}{\partial \gamma_{1}} &=  f(k, \tau) P_{1s}(k) (\hat{k}^{2}_{1} - \hat{k}^{2}_{2}) 
\end{align}
Expanding the expression in parentheses to first order in estimates $\hat \gamma_{1,2}$ and limiting ourselves to quadratic estimators, we find
\begin{equation}
	\hat \gamma_{1} = \int \frac{d^{3} k}{(2\pi)^{3}} \frac{|\tilde{\delta}(\textbf{k})|^{2}}{L^{3}} \frac{P_{1s}(k)}{\tilde{P}^2_{tot}(\textbf{k})} f(k, \tau) (\hat{k}^{2}_{1} - \hat{k}^{2}_{2}).
\end{equation}
We implicitly absorbed the normalization coefficient into the Wiener filter. Furthermore, we can rewrite this expression by first defining
\begin{equation}
	\tilde{\delta}^{w_{i}}(\textbf{k}) = \tilde{\delta}(\textbf{k})\left[ \frac{P_{1s}(k)f(k)}{\tilde{P}_{tot}^{2}(\textbf{k})} \right]^{1/2} i \hat{k}_{i}\label{eq:delta_gwi},
\end{equation}
and then inverse Fourier transforming $\tilde{\delta}^{w_{i}}$. Integral over $k$ yields a Dirac delta function cancelling the other position variable.
\begin{equation}
	\hat \gamma_1 = \int \frac{d^{3} x}{L^{3}} \left(\tilde{\delta}^{w_x}(\textbf{x})\tilde{\delta}^{w_x}(\textbf{x}) - \tilde{\delta}^{w_y}(\textbf{x})\tilde{\delta}^{w_y}(\textbf{x})\right) \label{eq:gamma1_integralestimate}
\end{equation}

We have defined the tidal field as the second derivatives of the gravitational potential at an observer, so the tidal field and consequently $\hat \gamma_i$ are not functions of \textbf{x}. Nevertheless, the estimate $\hat \gamma_1$ is an average over local values as equation~(\ref{eq:gamma1_integralestimate}) indicates. Therefore, the estimate for $\gamma_1$ at a location $\textbf{x}$ is the expression in parentheses. These estimators are effectively weighted derivatives. Following the same steps, we find an expression for $\hat \gamma_2$.
\begin{align}
	\hat \gamma_{1}(\textbf{x}) &= \delta^{w_{x}}(\textbf{x})\delta^{w_{x}}(\textbf{x}) - \delta^{w_{y}}(\textbf{x})\delta^{w_{y}}(\textbf{x}) \label{eq:gamma1_x}\\
	\hat \gamma_{2}(\textbf{x}) &= \delta^{w_{x}}(\textbf{x})\delta^{w_{y}}(\textbf{x}) + \delta^{w_{y}}(\textbf{x})\delta^{w_{x}}(\textbf{x}) \label{eq:gamma2_x}
\end{align}

Underlying density fluctuations and $\gamma_i$ are theoretically related through $k^2 \gamma_1 \propto (k_x^2-k_y^2)\delta_L/2$ and $k^2\gamma_2 \propto k_xk_y\delta_L$, so either $\gamma_i$ can be used to estimate $\delta_L$. Given $\gamma_1$ vanishes when $k_x=k_y$ and $\gamma_2$ vanishes when either $k_x$ or $k_y$ is zero, a better estimate for $\delta_L$ is the three dimensional convergence field \citep{kaiser_weak_1992}.
\begin{equation}
	\kappa_{3D}(\textbf{k}) = \frac{2k^{2}}{3(k_{x}^{2} + k_{y}^{2})^{2}} \left[ (k_{x}^{2} - k_{y}^{2}) \hat \gamma_{1}(\textbf{k}) + 2k_{x}k_{y} \hat \gamma_{2}(\textbf{k}) \right] \label{eq:kappa3d}
\end{equation}

However, we still need to filter out noise and correct for multiplicative biases in $\kappa_{3D}$. The Wiener filter is constructed by minimizing the error $e= \langle (W \kappa_{3D} - \delta)^2 \rangle$ with respect to $W$, which yields
\begin{equation}
    W(k_{\bot}, k_{\|}) = \frac{\langle \kappa_{3D} \delta \rangle}{\langle \kappa_{3D} \kappa_{3D} \rangle}.
\end{equation}
We compute these expectation values by direct simulations, although we also present an analytic approach in Section~\ref{sec:reconst_theory}.



\section{Implementation}
\label{sec:implementation}
\subsection{Simulations}
\label{subsec:simulations}
We run 10 simulations with $1024^{3}$ dark matter particles using GADGET-2\footnote{\url{http://wwwmpa.mpa-garching.mpg.de/gadget/}} \citep{springel_gadget:_2001,springel_cosmological_2005} in a box of side length $L = 1.5$ Gpc$/h$ with periodic boundary conditions. The boxes have the following cosmological parameters: $\Omega_{m} = 0.276$, $\Omega_{b} = 0.045$, $\Omega_{\Lambda} = 0.724$, $h = 0.7$, $n_{s} = 0.961$ and $\sigma_{8} = 0.811$. The simulations start from $z = 49$ using 2LPT initial conditions \citep{scoccimarro_transients_1998,jenkins_second-order_2010} constructed with the linear power spectrum from CAMB\footnote{\url{https://lambda.gsfc.nasa.gov/toolbox/tb_camb_form.cfm}}.

In the following sections, we compare the reconstruction results at $z=0$ and $z=1$ in real and redshift space without any foreground subtraction. For every other case, we run the reconstruction only at $z=1$.

\subsection{Method}
\label{subsec:method}
The tidal reconstruction algorithm presented in \citet{zhu_cosmic_2016} consists of two phases. The first phase computes $\kappa_{3D}$ from noisy over-density field, while the second phase applies a Wiener filter to $\kappa_{3D}$ to obtain the clean reconstructed field $\kappa$. 

In the first phase: 
\begin{enumerate}
    \item \label{it:step1} We interpolate the density field to a grid and smooth it with a Gaussian window function.
    
    \item \label{it:step2} We apply Gaussian mapping\footnote{Even though the log transform is a fast and simple Gaussianization procedure, the foreground subtraction models will cause $\delta < -1$ in our simulations, which hinders evaluating $\ln(1+\delta)$. Also considering the possible numerical errors at void regions, we have chosen to map the field into a Gaussian distribution by preserving the ranking. Given that any interferometic map will have regions $\delta < -1$ with or without foreground subtraction, this mapping will be required in real data as well.} while keeping the standard deviation same ($\sigma(\delta_G) = \sigma(\delta_R)$) to obtain the Gaussianized over-density field $\delta_{G}(\textbf{x})$ \citep{weinberg_reconstructing_1992}.
    
    \item \label{it:step3} We construct $\delta_{G}^{w_{i}}(\textbf{x})$ using equation~(\ref{eq:delta_gwi}).
    
    \item \label{it:step4} We estimate $\gamma$'s from equations~(\ref{eq:gamma1_x}) and~(\ref{eq:gamma2_x}).
    
    \item \label{it:step5} We use equation~(\ref{eq:kappa3d}) to determine the three dimensional convergence field $\kappa_{3D}$.
\end{enumerate}
Note that \ref{it:step3} and \ref{it:step5} are done in Fourier space, while \ref{it:step4} is a configuration space operation. As is normal, all smoothing steps are done in Fourier space. We use FFTW\footnote{\url{http://www.fftw.org}} for Fourier transforms on $1536^3$ grids and $R = 1.5$ Mpc/h as our smoothing radius, enough to suppress the interpolation kernel.

To estimate $\delta_{G}^{w_{i}}$, we adopt $P_{1s}(k)$ as the linear power spectrum from CAMB and non-linear power spectrum for $\tilde{P}_{tot}(\textbf{k})$. In redshift space we use a simple Kaiser form: $\tilde{P}^{(s)}_{tot}(\textbf{k}) = (1+f \mu^{2})^{2} \tilde{P}_{tot}(\textbf{k})$. The reconstruction works on $xy$ slices and does not have information in purely $z$ direction. As a result $\kappa_{3D}(k_{\bot} = 0, k_z)$ modes go to infinity and contain only noise. We set these modes to zero. Since peculiar velocities cause distortions in $x_{\|}$ direction, we expect $\gamma_{1,2}$ not to be particularly affected by redshift space distortions. 

$\kappa_{3D}$ is a biased estimate of $\kappa$; it is noisy and missing a normalization constant. We construct Wiener filter from ten $\langle \kappa_{3D}\delta \rangle$ and $\langle\kappa_{3D}\kappa_{3D}\rangle$ pairs. To get better statistics, we bin the spectra in $(k_{\bot}, k_{\|})$. We linearly interpolate the Wiener filter in 2D using GSL\footnote{\url{https://www.gnu.org/software/gsl/}}. In the second phase, the reconstructed field $\kappa$ is the filtered three dimensional convergence field.
\begin{equation}
	\kappa(\textbf{k}) = \kappa_{3D}(\textbf{k})W(k_{\bot}, k_{\|})
\end{equation}
We stress the ensemble (and angle) average in the Wiener filter expression. This filtering does not perfectly recover the true over-density field.

\begin{figure}
	\includegraphics[width=\columnwidth]{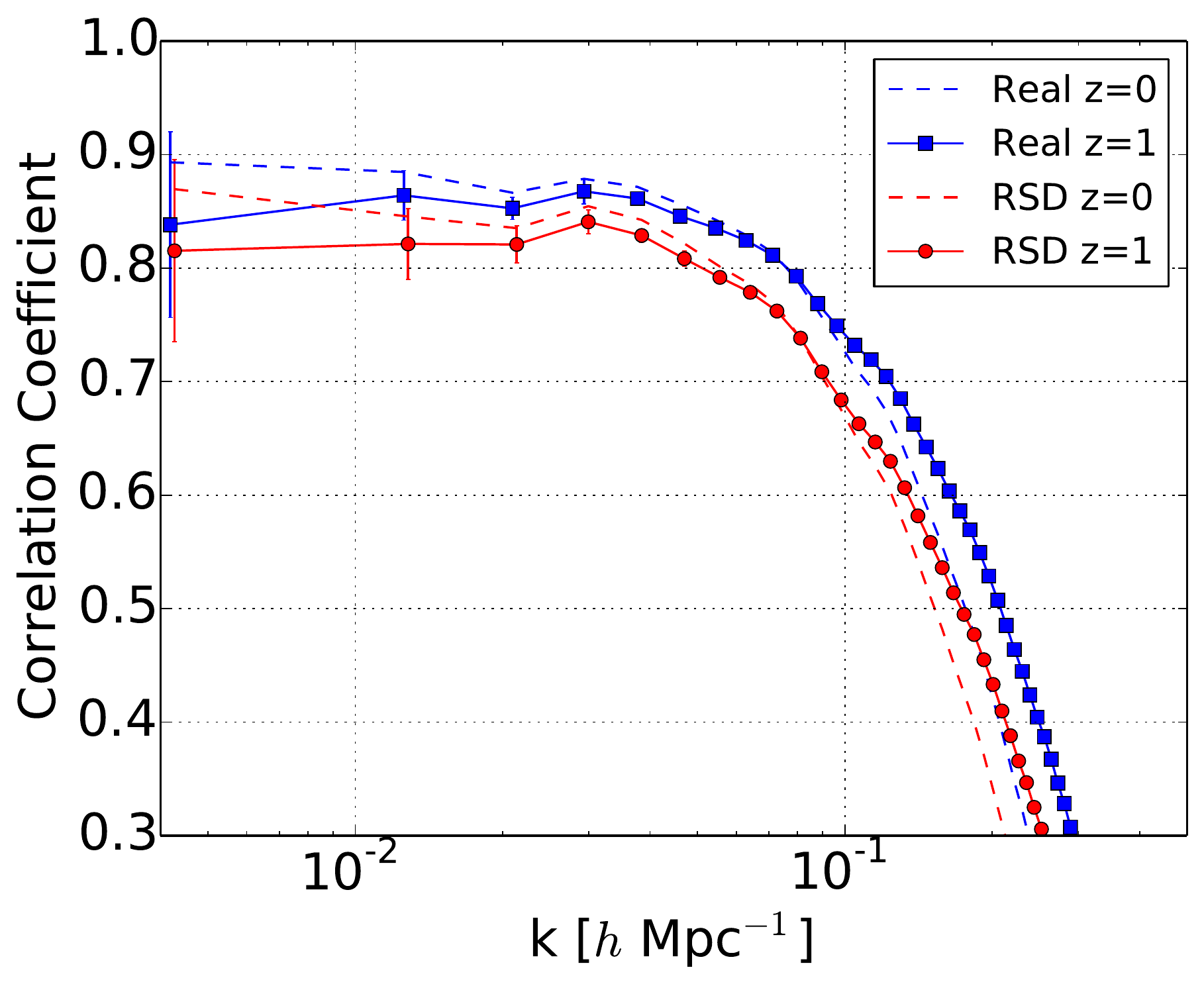}
	\includegraphics[width=\columnwidth]{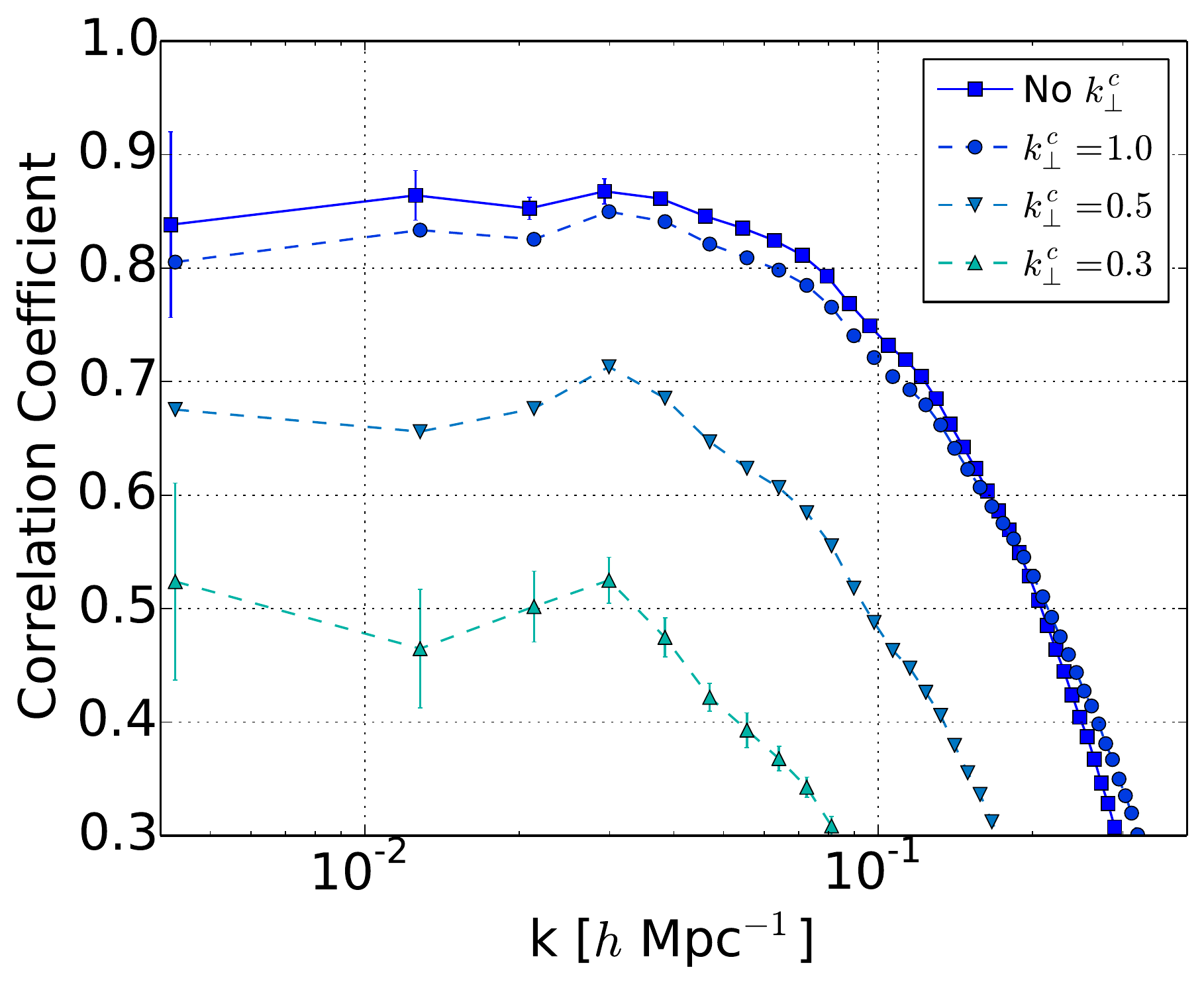}
	\caption{(Top) The cross-correlation coefficients $r_{\kappa\delta}(k)$ in real and redshift spaces without removing any modes at $z=0$ and $z=1$. The redshift space distortions cause $\approx 0.03$ decrease in the correlation coefficient. (Bottom) $r_{\kappa\delta}(k)$ after applying $k_{\bot}$ cutoff at redshift $z=1$ in real space with $R=1.5$ Mpc$/h$. Losing small-scale angular modes decreases the correlation coefficient.}
	\label{figure:compare_real_rsd_nofg_compare_kperp_real_z1}
\end{figure}



\section{Results}
\label{sec:results}
Throughout this paper we plot the mean cross-correlation coefficients from N-body simulations between true over-density field $\delta$ and the reconstructed field $\kappa$, where $r_{\kappa\delta} = P_{\kappa\delta}/\sqrt{P_{\delta}P_{\kappa}}$. Error bars are computed from the standard deviation of ten simulation results. 

\begin{figure}
	\includegraphics[width=\columnwidth]{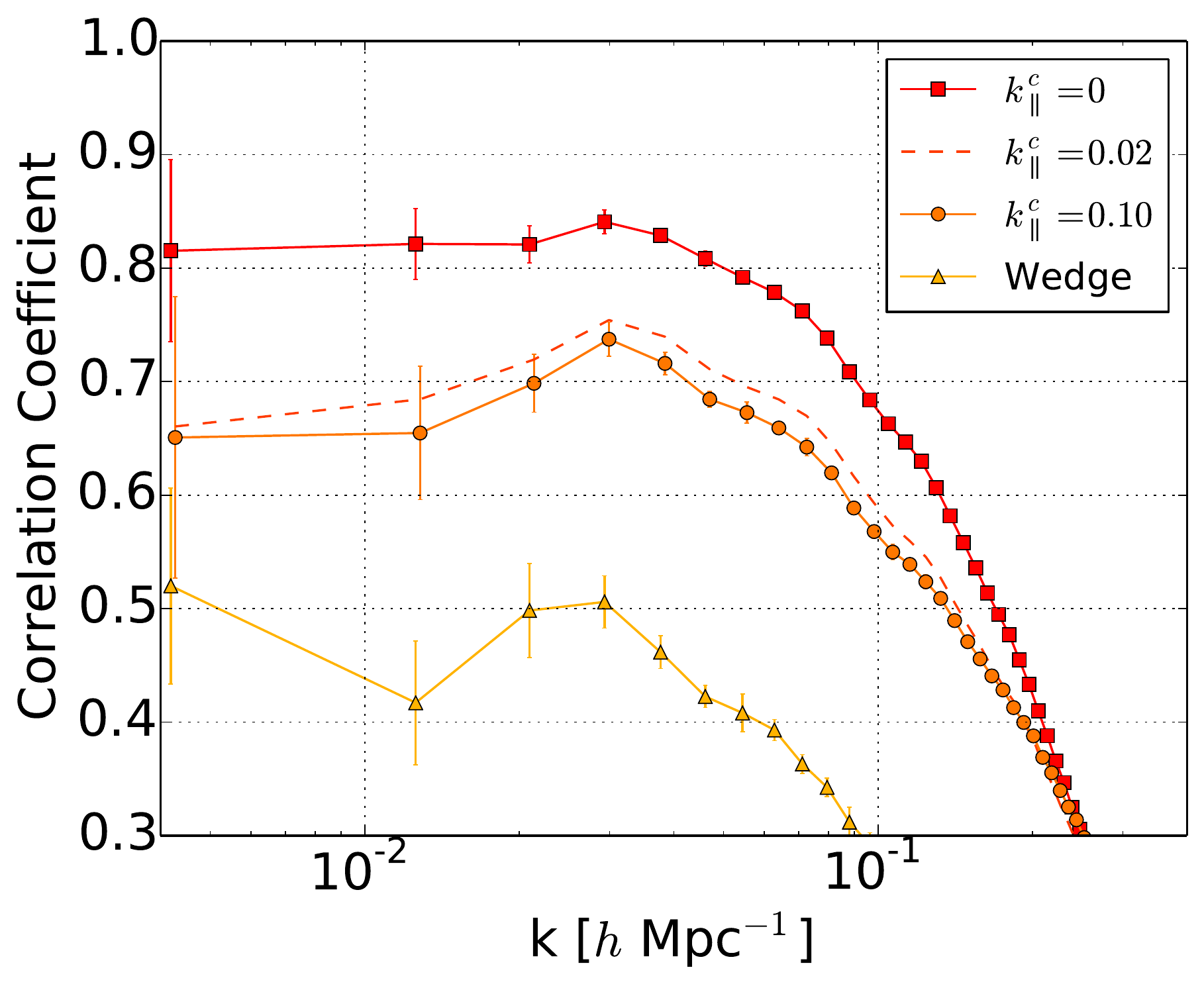}
	\caption{The cross-correlation coefficients $r_{\kappa\delta}(k)$ at $z=1$ with $R=1.5$ Mpc$/h$ in redshift space after applying the low $k_{\|}$ cutoff and removing foreground wedge. After decreasing approximately $0.15$, the correlation coefficient is not sensitive to $k^c_{\|}$ up to $0.1$ $h/$Mpc. Removing the wedge modes impairs the reconstruction and makes the $k^c_{\|}$ value irrelevant.}
	\label{figure:compare_lowkpara_all_z1}
\end{figure}

\begin{figure*}
	\includegraphics[width=0.325
    \linewidth]{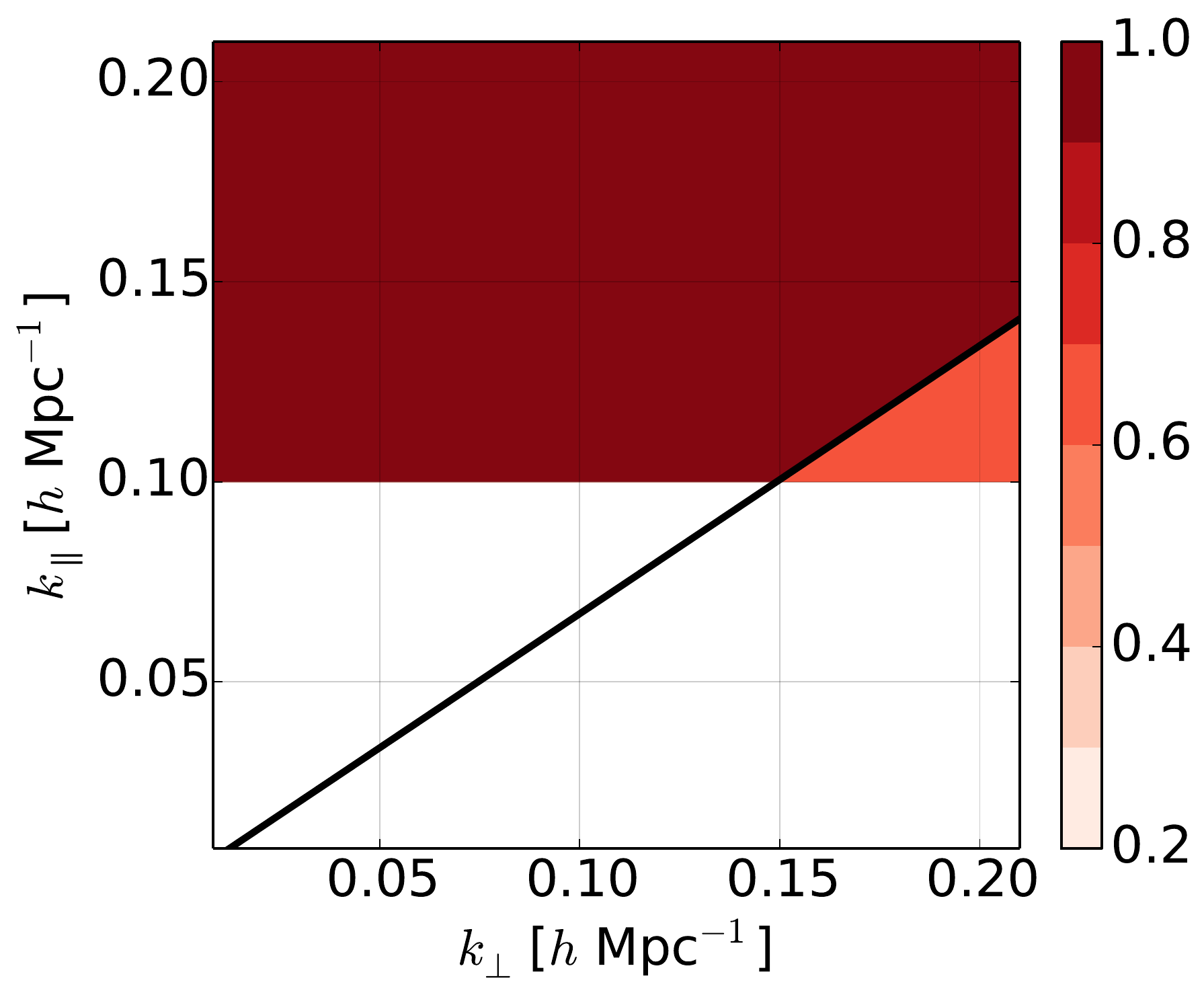}
    \includegraphics[width=0.325
    \linewidth]{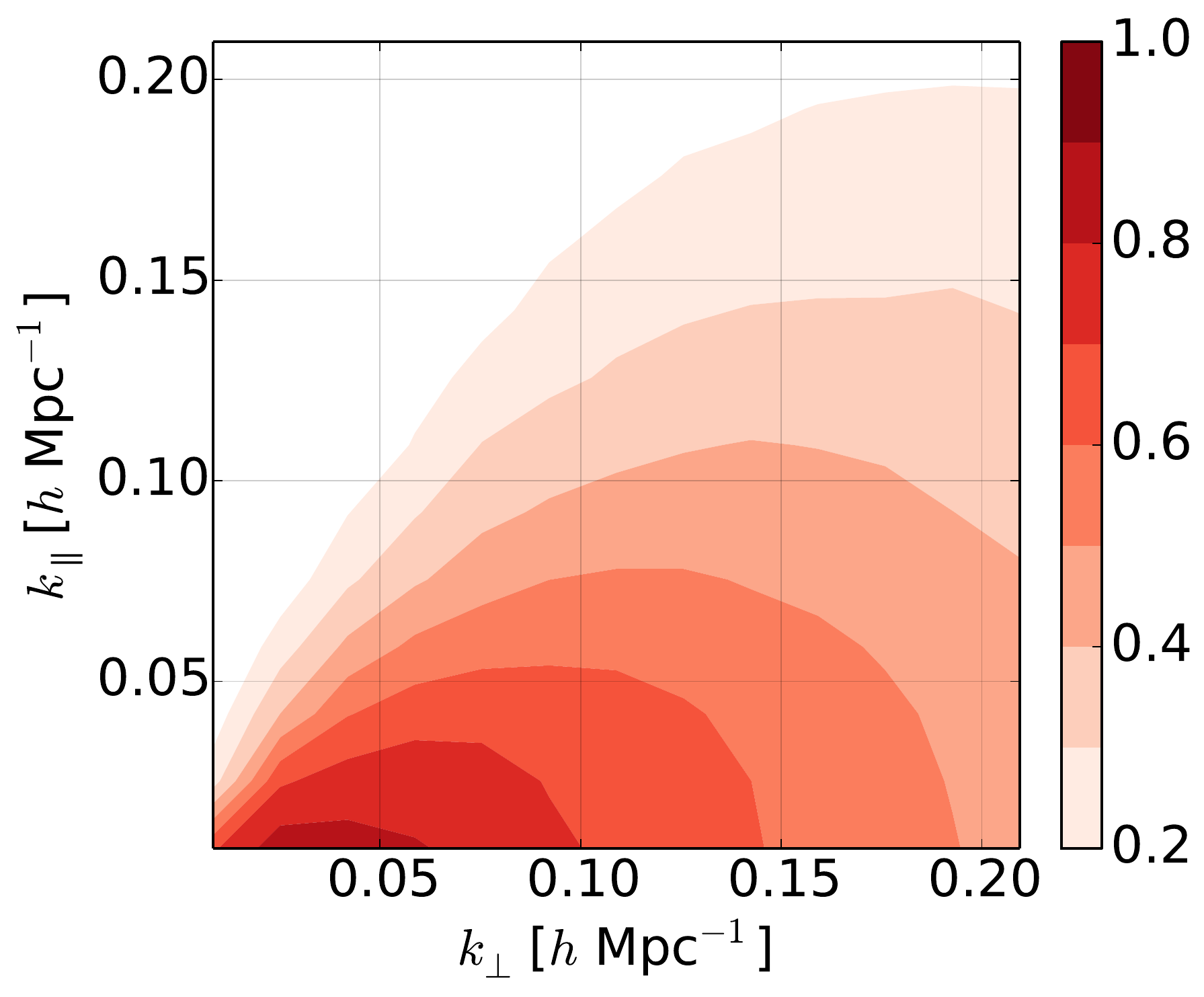}
    \includegraphics[width=0.325
    \linewidth]{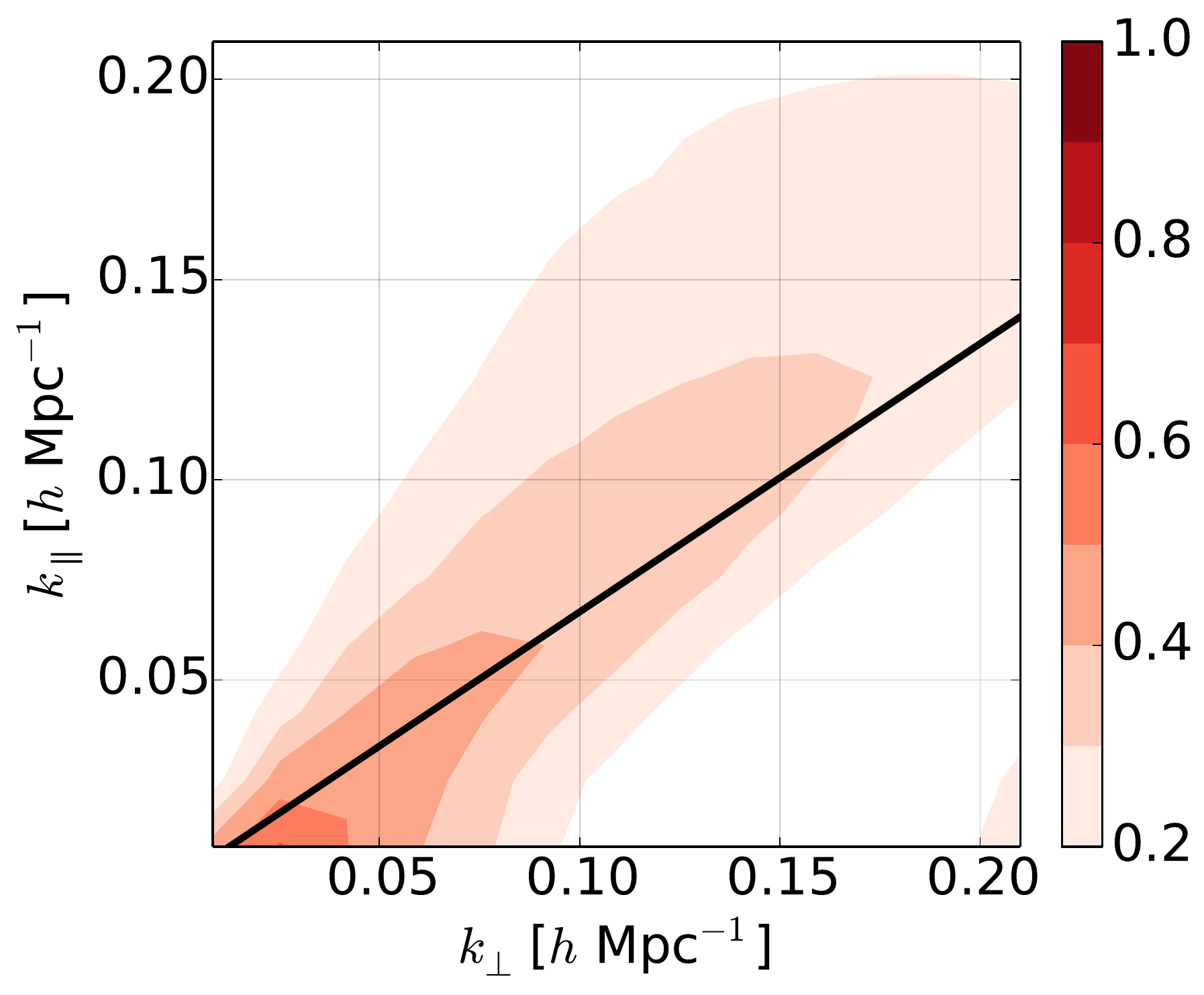}
    \caption{The cross-correlation coefficients $r(k_{\bot}, k_{\|})$ at $z=1$ with $k^c_{\|}=0.1$ $h/$Mpc in redshift space. (Left) A cartoon for foreground removal which does not include redshift space distortions. Removed wedge modes are represented by bright red colour. They also include high $k$. (Middle) If only low $k_{\|}$ modes are removed, the reconstruction recovers correlations up to $k_{\bot} \approx 0.2$ $h/$Mpc. As expected, low $k_{\bot}$ with $k_{\|} \gtrsim 0.02$ $h/$Mpc modes contain noise and are not correlated with the original over-density field. (Right) Removing wedge modes reduces $r(k_{\bot}, k_{\|})$ and limits the reconstruction to a triangle up to $k_{\bot} \approx 0.05$ $h/$Mpc with low $k_{\|}$.}
    \label{figure:FG_Kpara010_foreground_rec_z1_aniso_ccc}
\end{figure*}

\begin{figure}
    \centering
    \includegraphics[width=\columnwidth]{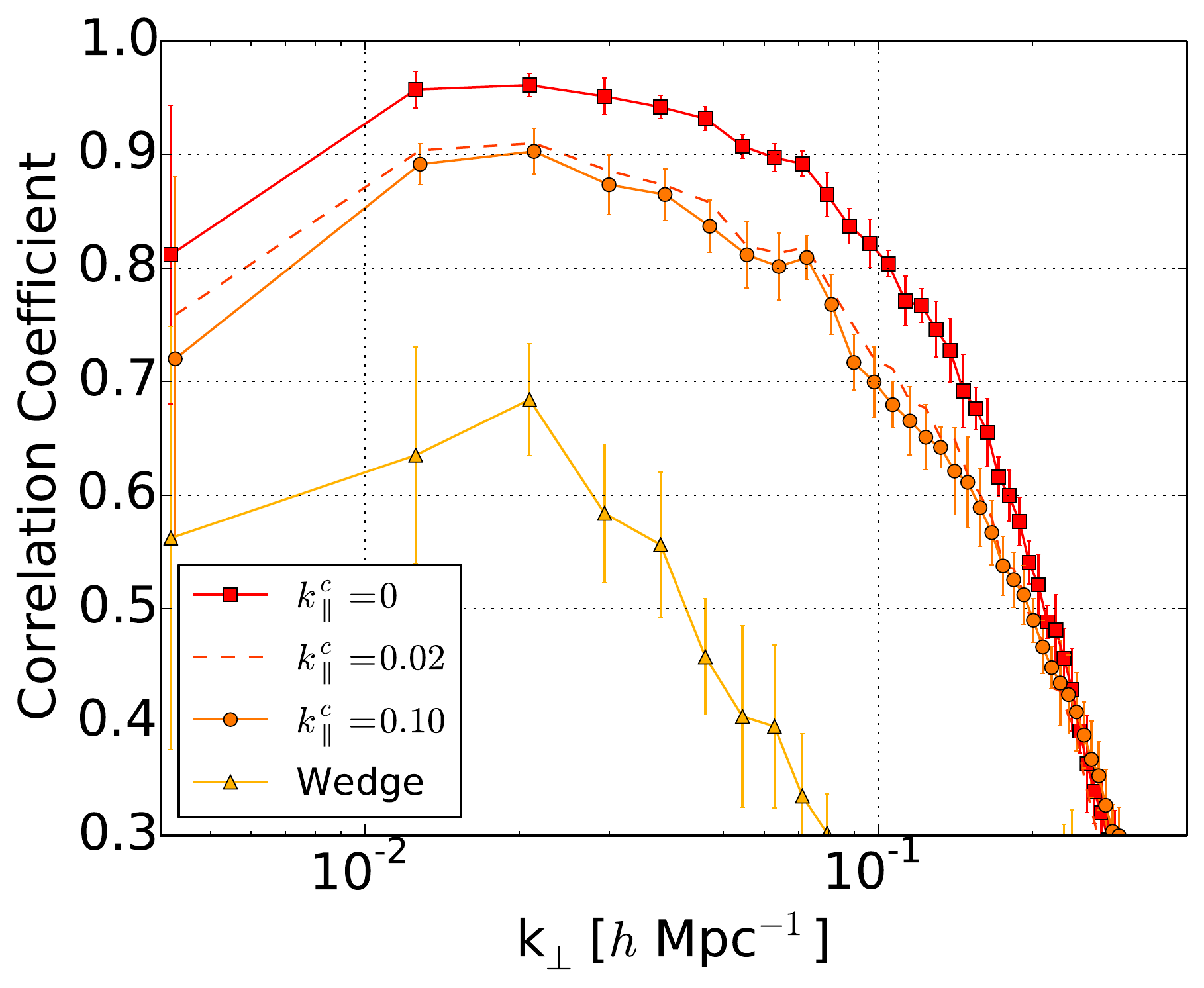}
    \caption{The same figure as Fig.~\ref{figure:compare_lowkpara_all_z1}, but for cross correlation coefficients $r_{\kappa\delta}(k_\bot, k_\|=0)$. Even though the wedge case is still in poor quality, the reconstruction does significantly better on this plane.}
    \label{fig:kz0_iso_compare}
\end{figure}

\subsection{Tests on the Reconstruction}
We run the reconstruction on full dark matter field in real and redshift spaces at $z=0$ and $z=1$. On the top panel of Fig.~\ref{figure:compare_real_rsd_nofg_compare_kperp_real_z1}, solid lines represents the results for $z=1$, whereas dashed lines represents the results for $z=0$. The results are similar for all cases, so we explicitly show data points and error bars only for $z=1$. The redshift space distortions (red circles) degrade the correlation coefficient by less than 5\%. Given the reconstruction relies on non-linear coupling, its efficiency depends on redshift in a complicated way. For example, purely linear regime ($z\rightarrow\infty$) has no coupling to exploit, whereas late times has higher order coupling besides nearest order. Nevertheless, we do not see a strong improvement at higher redshift. Overall, the reconstruction keeps the correlations high ($\gtrsim 0.7$) up to $k\approx 0.1$ $h/$Mpc. However, our results are approximately $10\%$ worse than the results of \citet{zhu_cosmic_2016}. We attribute this difference to a combination of our lower particle density and higher smoothing radius.

The reconstruction is based on non-linear tidal interactions, so it needs small-scale modes to work efficiently. These modes are limited by the finite beam size, foreground wedge effect and the smoothing radius. Increasing the smoothing radius, we confirm the findings of \citeauthor{zhu_cosmic_2016}: The smoothing radius $R$ should not be larger than 5 Mpc$/h$. To simulate the finite beam, we remove high $k_{\bot}$ modes since they cannot be well resolved. We implement these as in the foreground case described below. Note that this cutoff makes small-scale signal weaker, but unlike Gaussian smoothing it is a real loss of information and not isotropic. The bottom panel of Fig.~\ref{figure:compare_real_rsd_nofg_compare_kperp_real_z1} demonstrates the reconstruction mostly uses modes with $k_{\bot} \la 1.0$ $h/$Mpc and does not work well without modes with $k_{\bot} \ga 0.3$ $h/$Mpc. Results in the redshift space are similar and are not depicted. In reality, how well an angular scale can be resolved depends on the baseline distribution for the experiment. A HIRAX-like experiment has access to a broad range of angular scales; $k_{\bot} < 1$ $h/$Mpc are present, although with diminishing quantity \citep[see][figure 1]{white_matched_2017}. Moreover, the thermal noise power is a function of $k_{\bot}$. Simple estimates yield values between 150-600 Mpc$^3/h^3$ at $z=1$ and $k_{\bot}=0.2$ $h/$Mpc \citep[see][Appendix A]{white_matched_2017} and cross over at $k=0.6$ $h/$Mpc \citep{zhu_recovering_2018}. Since our figures and results show simply removing modes, the reality will be between the cases we have presented.

\subsection{Foregrounds}
\label{subsec:foregrounds}
In this section, we investigate if the reconstruction can recover absent modes. Foregrounds from galactic and extra-galactic synchrotron and free-free emissions are spectrally smooth and contaminate modes with small $k_{\|}$. Without any angular resolution cutoff, we now turn to the astrophysical foregrounds problem at hand. These foregrounds have been implemented using a high-pass filter in \citet{zhu_recovering_2018}. However, estimator construction in Section~\ref{subsec:estimator} has motivated us to take a different approach. Instead of a high-pass filter, we adopt a sharp cutoff in $k_{\|}$ and assume $P_N(\textbf{k})$ to be infinity for every mode set to zero. We first remove $k_z < k_{\|}^{c}$ modes from the original over-density field. Since these modes are purely noise, we set $\delta_{g}^{w_{i}}(k_{z} < k_{\|}^{c})$ in equation~(\ref{eq:delta_gwi}) to zero explicitly. The reconstruction recovers discarded modes in configuration space. We stress that these foregrounds cost high $k_\bot$ modes as well and reduce the reconstruction's efficiency.

Instrumental limitations introduce another challenge by allowing low $k_{\|}$ foregrounds to leak into higher $k_{\bot}-k_{\|}$ modes \citep{morales_four_2012}. This effect forms a foreground wedge and contaminates modes with $|k_{\|}| \leq m(z) k_{\bot}$, where $m(z) = D_M(z)H(z)/(1+z)$ and $D_M(z)$ is the co-moving angular diameter distance \citep{seo_foreground_2016, white_matched_2017}. We set these modes to zero to realize the foreground wedge. This effect costs even more small-scale modes and significantly limits the reconstruction's efficiency.

We remind the reader of our two-phase procedure. The first phase uses the foreground subtracted over-density field. The second phase reverts to using the true over-density field $\delta$ to construct the Wiener filter. 

We test three values for $k^c_{\|}$ ($0.02$ $h/$Mpc, $0.05$ $h/$Mpc and $0.1$ $h/$Mpc) in real and redshift space at $z=1$. The correlation coefficient falls sharply after the foreground removal since we remove high $k_{\bot}$ modes in the process. However, we find the results are not truly sensitive to $k^c_{\|}$. The reconstruction can recover lost modes  up to $k^c_{\|} = 0.1$ $h/$Mpc. Fig.~\ref{figure:compare_lowkpara_all_z1} displays the results in redshift space. Real space results follow the same trend with a better performance  similar to Fig.~\ref{figure:compare_real_rsd_nofg_compare_kperp_real_z1}. 

The reconstruction deteriorates more dramatically in the presence of a foreground wedge as seen in Fig.~\ref{figure:compare_lowkpara_all_z1}. We show the results for $k^c_{\|} = 0.1$ $h/$Mpc; changing $k^c_{\|}$ has no effect anymore since the additional information in low $k_{\bot}$ triangle is insignificant. 

In Fig.~\ref{figure:FG_Kpara010_foreground_rec_z1_aniso_ccc}, we show the cross correlation coefficient in the anisotropic $(k_{\bot}, k_{\|})$ plane. The figure on the left is a cartoon for the foreground removal procedure, which does not depict redshift space distortions; and the removed wedge modes are represented in bright red. The middle panel shows the results for the low $k_{\|}$ removal. Modes with $k_{\|} < 0.1$ $h/$Mpc are recovered up to $k_{\bot} \approx 0.2$ $h/$Mpc, but the anisotropic nature of reconstruction causes a lower $r(k)$ at $k \approx 0.2$ $h/$Mpc in Fig.~\ref{figure:compare_lowkpara_all_z1}. Without the wedge modes the reconstruction cannot recover $k_{\bot} \ga 0.1$ $h/$Mpc modes, and its performance is limited around the wedge line with $k_{\bot} \la 0.05$ $h/$Mpc. The performance of the reconstruction on $k_\|=0$ plane is shown in Fig.~\ref{fig:kz0_iso_compare}. The correlations are significantly higher on this plane as Fig.~\ref{figure:FG_Kpara010_foreground_rec_z1_aniso_ccc} signifies. These modes weigh more in weak lensing cross correlations.

The level to which the foreground wedge can be corrected will depend on the experiment and is still unknown. So, we prefer isolating the wedge effect from low angular resolution, whereas in reality both effects will be present. The details of foregrounds and available angular modes will depend on the instrument and should be calculated into $P_N$ for an exact treatment. Although it would deteriorate the reconstruction further, combining these two does not necessarily negate the recovery, since there is an overlap between two cases.



\section{Theoretical Expectations}
\label{sec:reconst_theory}
The cosmic tidal reconstruction combines distinct theoretical pieces such as tidal interactions, maximum likelihood estimators and three dimensional convergence field, then adds another complication by requiring multiple Fourier transforms (see Section~\ref{subsec:method} for details). Due to these difficulties, previous works relied only on simulations to examine the reconstruction. Incorporating all the theoretical segments into one framework, we derive analytic expressions for cross and power spectrum. These expressions yield the correct structure in power spectrum and quantify the expectations in reconstruction efficiency due to lost modes and peculiar velocities. This framework also promises an estimate for Wiener filter when higher order terms are considered.

We start by defining the Gaussianized over-density field $\delta_{G}(\textbf{x}) = f_{G}[\delta_{R}(\textbf{x}); \sigma]$, where $\delta_R$ is smoothed with a Gaussian window function with radius $R$ and $f_G$ is the Gaussian mapping. Then, we rewrite equation~(\ref{eq:delta_gwi}) as $\delta_{G}^{w_{i}}(\textbf{k}) = \delta_{G}(\textbf{k}) w(\textbf{k}) (ik_{i}/k)$. We transform equations~(\ref{eq:gamma1_x}) and~(\ref{eq:gamma2_x}) for $\gamma_i(\textbf{x})$ into $k$ integrals by substituting Fourier transforms of $\delta_{G}^{w_{i}}(\textbf{x})$.
\begin{align}
	\gamma_{i}(\textbf{x}) &=  \int \frac{d^{3}k d^{3}q}{(2\pi)^{6}} e^{i (\textbf{k} +\textbf{q})\cdot \textbf{x}} \tilde{T}_{\gamma_{i}}(\textbf{k}, \textbf{q}) \delta_{G}(\textbf{k})\delta_{G}(\textbf{q}),
\end{align}
where
\begin{align}
	\tilde{T}_{\gamma_{1}}(\textbf{k}, \textbf{q}) &= \left[ - \frac{k_{x}q_{x} -k_{y}q_{y} }{k q} \right]w(\textbf{k})w(\textbf{q}) \\
	\tilde{T}_{\gamma_{2}}(\textbf{k}, \textbf{q}) &=\left[ - \frac{k_{x}q_{y}  + k_{y}q_{x} }{k q} \right]w(\textbf{k})w(\textbf{q}).
\end{align}
We warn the reader that $\textbf{q}$ in this context does not represent Lagrangian coordinates.

Note that $w(\textbf{k}) = w(-\textbf{k})$ and therefore $T_{\gamma_{i}}$ is symmetric and has the following properties:
\begin{align}
	\tilde{T}_{\gamma_{i}}(\textbf{k}, \textbf{k}^{\prime}) &= \tilde{T}_{\gamma_{i}}(\textbf{k}^{\prime}, \textbf{k}) = \tilde{T}_{\gamma_{i}}(-\textbf{k}, -\textbf{k}^{\prime}) \\
	&= -\tilde{T}_{\gamma_{i}}(\textbf{k}, - \textbf{k}^{\prime}) = -\tilde{T}_{\gamma_{i}}(-\textbf{k}, \textbf{k}^{\prime}).
\end{align}

Now, we find the Fourier transforms of $\gamma$'s.
\begin{align}
	\gamma_{i}(\textbf{k}) &= \int d^{3}x e^{-i\textbf{k}\cdot\textbf{x}} \gamma_{i}(\textbf{x}) \\
	&= \int d^{3}x \frac{d^{3}q d^{3}p}{(2\pi)^{6}} e^{i (\textbf{q} + \textbf{p} - \textbf{k}) \cdot \textbf{x}} \tilde{T}_{\gamma_{i}}(\textbf{q}, \textbf{p}) \delta_{G}(\textbf{q})\delta_{G}(\textbf{p})
\end{align}
The integral over $x$ yields a Dirac delta function which integrates out $p$. Consequently, the local estimates for each $\gamma$ becomes
\begin{equation}
	\gamma_{i}(\textbf{k})=  \int\frac{d^{3}q}{(2\pi)^{3}} \tilde{T}_{\gamma_{i}}(\textbf{q}, \textbf{k} - \textbf{q}) \delta_{G}(\textbf{q})\delta_{G}(\textbf{k} - \textbf{q}) \label{eq:th_gammai}
\end{equation}
in Fourier space. Equation~(\ref{eq:th_gammai}) easily extends to an expression for the three dimensional convergence field $\kappa_{3D}(\textbf{k})$. We prefer defining $\textbf{p} \equiv \textbf{k} - \textbf{q}$ to further simplify the notation. Using equation~(\ref{eq:kappa3d}), we arrive at
\begin{equation}
	\kappa_{3D}(\textbf{k}) = \int\frac{d^{3}q}{(2\pi)^{3}} \tilde{T}_{\kappa_{3D}}(\textbf{q}, \textbf{p}) \delta_{G}(\textbf{q})\delta_{G}(\textbf{p}) \label{eq:analytic_kappa3d},
\end{equation}
where $\tilde{T}_{\kappa_{3D}}(\textbf{q}, \textbf{p})$ is given by equation~(\ref{eq:kappa3d}) with $\tilde{T}_{\gamma_{i}}$ instead of $\gamma_i$. The exact form of $\tilde{T}_{\kappa_{3D}}$ can be found in Appendix \ref{app:reconst_theory}.

Our conventions for the cross spectrum and the power spectrum are the same: $\langle \kappa_{3D}(\textbf{k})\delta(\textbf{k}^{\prime}) \rangle = (2\pi)^3 \delta_D(\textbf{k} + \textbf{k}^{\prime}) P_{\kappa_{3D}\delta}(\textbf{k})$. Having a direct link between $\delta$ and $\kappa_{3D}$, we are now able to find these spectra. The cross spectrum is
\begin{equation}
	P_{\kappa_{3D}\delta}(\textbf{k}) =  \int\frac{d^{3}q}{(2\pi)^{3}}  B_T(\textbf{q}, \textbf{p})\tilde{T}_{\kappa_{3D}}( \textbf{q}, \textbf{p}), \label{eq:th_cross_kappa3d}
\end{equation}
where we have defined a modified bispectrum $B_T$ as
\begin{equation}
	\langle \delta_{G}(\textbf{q})\delta_{G}(\textbf{k} - \textbf{q})\delta(\textbf{k}^{\prime})\rangle = (2\pi)^{3}\delta_{D}(\textbf{k} + \textbf{k}^{\prime}) B_T(\textbf{q}, \textbf{p}).
\end{equation}

Although assuming $\delta_G$ as the linear field $\delta_l$ is tempting, this assumption results in a very small bispectrum in low $k$, since $\delta$ also approximates the linear field at these scales. Given $\delta_l$ is Gaussian and has zero bispectrum, $B_T \propto \langle \delta_l^3\rangle = 0$. From this short discussion, we conclude that second order corrections in $\delta_G$ have to be taken into account. This makes sense considering tidal reconstruction makes use of mode coupling terms which do not exist in linear theory. To find these necessary higher order contributions, we need to take a closer look at the Gaussianization procedure. Since the logarithmic transform is easier to handle \citep{neyrinck_rejuvenating_2011,mccullagh_recovering_2016}, whereas the exact Gaussian mapping is analytically complicated, we assume
\begin{align}
    \delta_G(\textbf{x}) &= \ln(1+\delta_R(\textbf{x})) - \langle \ln(1+\delta_R(\textbf{x})) \rangle \\
    &\approx  \delta_R - \frac{\delta_R^{2}}{2} + \frac{\delta_R^{3}}{3} - \frac{\delta_R^{4}}{4} + \cdots - \text{const}.
\end{align}
The constant average term is only in $\textbf{k} = 0$ mode and can be ignored. We ignore smoothing for simplicity, and refer the reader to Appendix \ref{app:smooth} for expressions with smoothing.  Taking the Fourier transform and using second-order perturbation results \citep{jain_second-order_1994}, we arrive at
\begin{equation}
	\delta_{G}(\textbf{k}) \approx \delta_{l}(\textbf{k}) + \int \frac{d^{3}q}{(2\pi)^{3}}  F_{2, G}(\textbf{q}, \textbf{p})  \delta_{l}(\textbf{q}) \delta_{l}(\textbf{p}),
\end{equation}
where $F_{2, G} = F_2 - \frac{1}{2}$ and
\begin{equation}
	F_{2}(\textbf{k}_1, \textbf{k}_2) = \frac{5}{7} + \frac{\textbf{k}_1 \cdot \textbf{k}_2}{2}\left[ \frac{1}{k_1^2} +\frac{1}{k_2^2} \right] + \frac{2}{7} \frac{(\textbf{k}_1\cdot \textbf{k}_2)^{2}}{k_1^{2}k_2^{2}}.
\end{equation}

Having these expressions, we can now calculate the modified bispectrum $B_T$. Its only difference from normal bispectrum $B$ calculation is $-1/2$ term in $F_{2, G}$, so this step is not too cumbersome.  Then,
\begin{equation}
	B_{T}(\textbf{q}, \textbf{p}) = B(\textbf{q}, \textbf{p}, -\textbf{k}) - P_{l}(k) \left(P_{l}(p)+ P_{l}(q)\right),
\end{equation}
where $P_l$ is the linear power spectrum and
\begin{equation}
    B(\textbf{k}_1, \textbf{k}_2, \textbf{k}_3) = 2 F_2(\textbf{k}_1, \textbf{k}_2) P_l(k_1) P_l(k_2) + \text{cyc.}
\end{equation}

A similar calculation shows that the four-point function gives the power spectrum of $\kappa_{3D}$. We remind the reader that the four-point function $\langle \delta^4 \rangle$ can be decomposed into three cyclic $\langle \delta^2\rangle^2$ terms and a connected term. The trispectrum is the Fourier transform of the connected four-point correlation function. In this estimation, assuming $\delta_G$ to be the linear field yields a zero trispectrum, but a non-zero four-point function with one vanishing cyclic term.
\begin{equation}
	P_{\kappa_{3D}} (\textbf{k}) = 2 \int \frac{d^{3}q}{(2\pi)^{3}} \tilde{T}^{2}_{\kappa_{3D}}(\textbf{q}, \textbf{p})  P_{l}(q) P_{l}(p) \label{eq:th_power_kappa3d}
\end{equation}
As we have discussed in the cross spectrum case, $\delta_G \approx \delta_l$ produces uncorrelated $\kappa_{3D}$. Hence, we would expect this equation to underestimate the true power.

With all limitations in mind, we can build a theoretical Wiener filter using equations~(\ref{eq:th_cross_kappa3d}) and~(\ref{eq:th_power_kappa3d}), and find the reconstructed power spectrum $P_{\kappa}$.
\begin{equation}
	P_{\kappa}(\textbf{k}) = \frac{P_{\kappa_{3D}\delta}^2}{P_{\kappa_{3D}}}
\end{equation}

\begin{figure}
    \centering
	\includegraphics[width=\columnwidth]{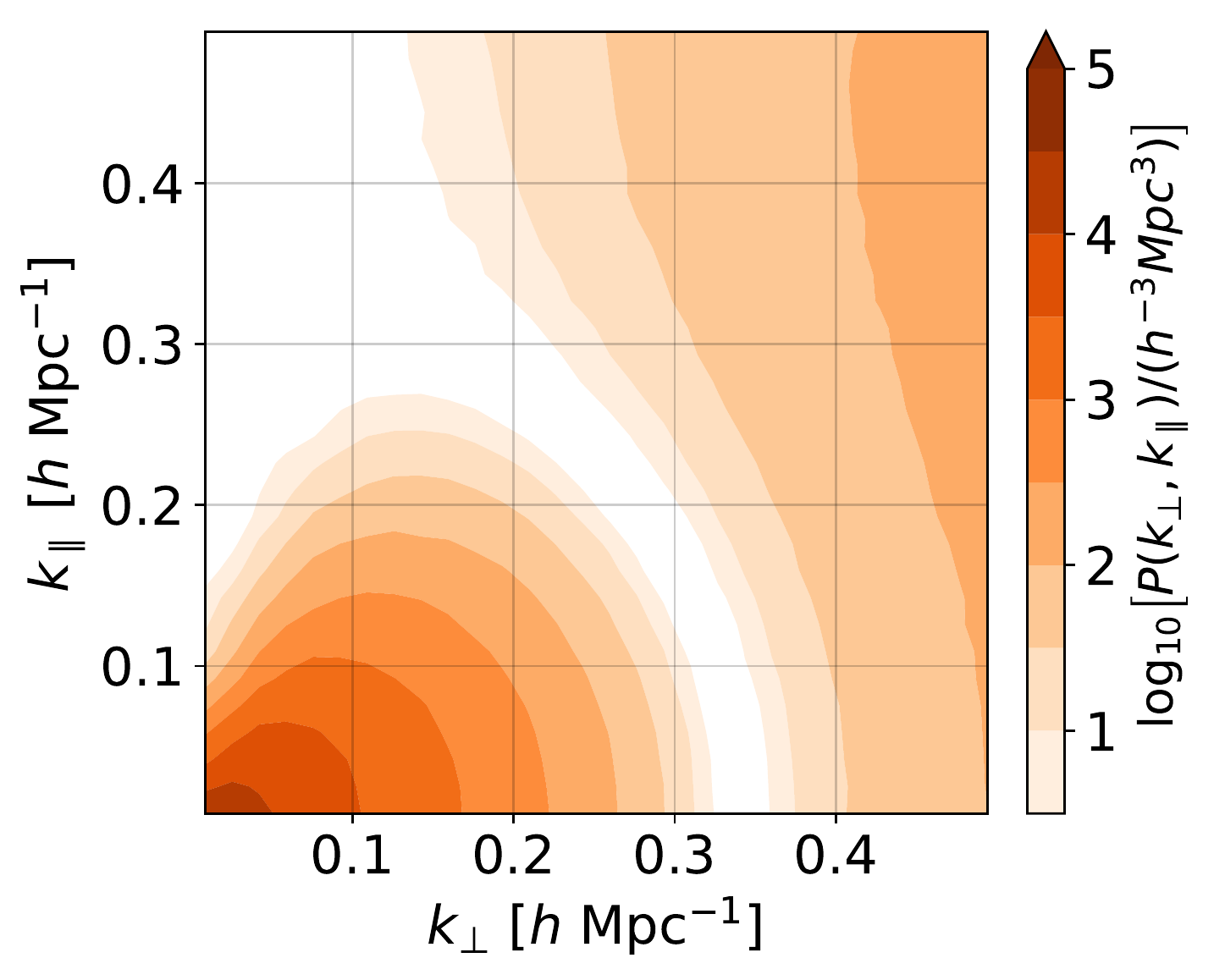}
	\includegraphics[width=\columnwidth]{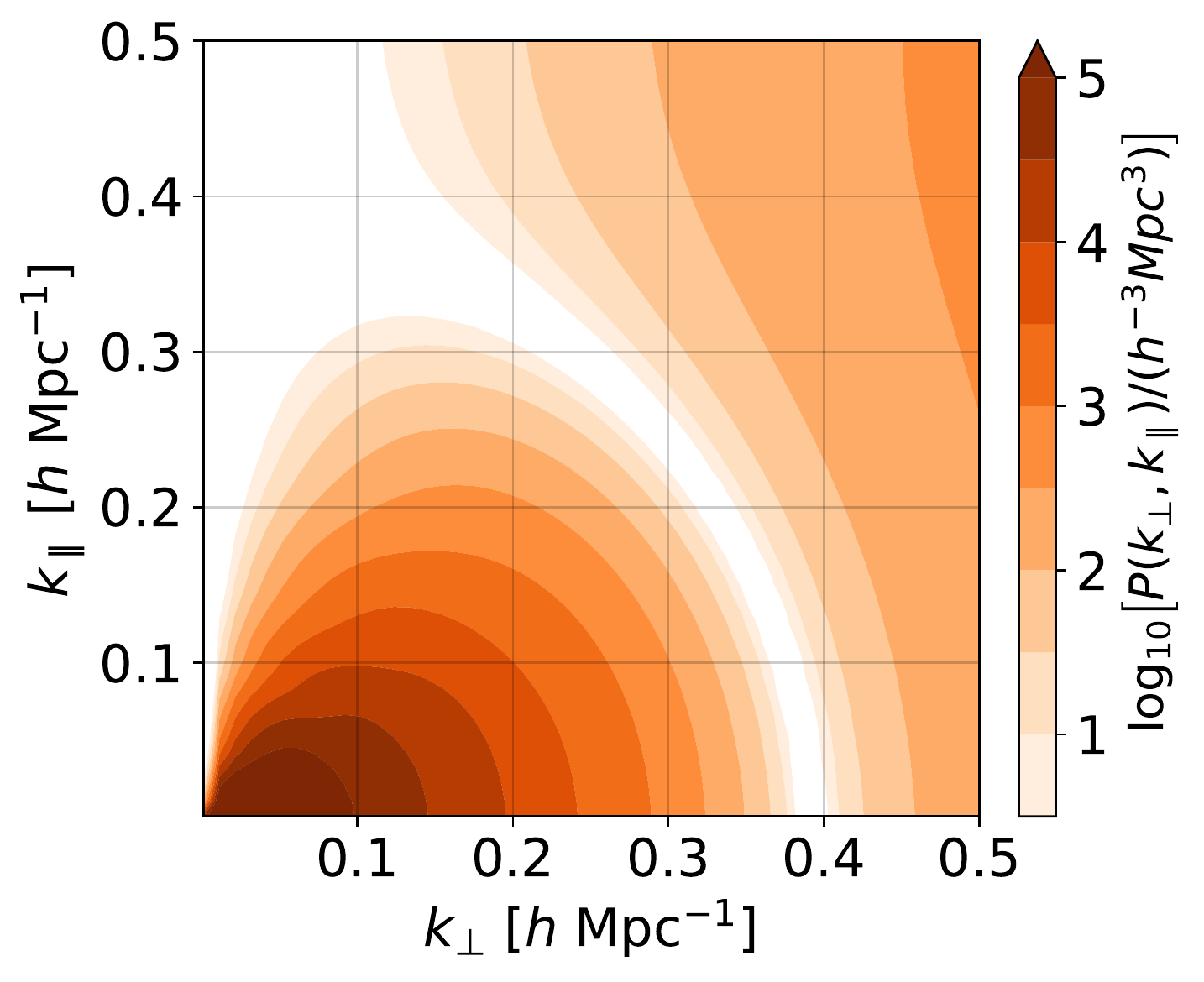}
	\caption{Power spectrum $P_{\kappa}$ of one simulation run (Top) compared to the theoretical estimate (Bottom). Even though theory correctly predicts the structure in the power spectrum, it yields larger $P_{\kappa}$ than numerical simulations. Our $P_{\kappa_{3D}}$ estimate is actually uncorrelated due to nearest order approximation, and lacks the reconstruction signal. This lack of power in $\kappa_{3D}$ results in high $P_{\kappa}$.}
	\label{figure:power_theory_vs_sim}
\end{figure}

\subsection{Results}
Having performed extensive numerical analyses in Section~\ref{sec:results}, we now test our theoretical framework to understand the emergent features and limited efficiency of cosmic tidal reconstruction. Our focus is on equations~(\ref{eq:th_cross_kappa3d}) and~(\ref{eq:th_power_kappa3d}). We include smoothing in our calculations (see Appendix \ref{app:smooth}) and take $R=1.5$ Mpc$/h$ as before.

The volume integration is simpler in spherical coordinates.
\begin{equation}
	\int d^3q \rightarrow \int_{q_i}^{q_f} q^2 dq \int_{-1}^1 dv \int_{-\pi}^{\pi} d\phi,
\end{equation}
where $\theta$ is the azimuthal angle with respect to the line-of-sight and $v=\cos\theta$. We apply two more transformations. First, we write the $\phi$ integral in terms of $t=\cos\phi$, since this integrand depends only on $\cos\phi$ and it is symmetric under $\phi \rightarrow -\phi$. Second, we use logarithmic spacing in $q$ integration. Putting these together, the volume integration becomes
\begin{equation}
	\int d^3q \rightarrow \int_{\ln q_i}^{\ln q_f} q^3 d\ln q \int_{-1}^1 dv \int_{-1}^{1} \frac{2dt}{\sqrt{1-t^2}}.
\end{equation}

We use GSL's Monte-Carlo integration library to compute these integrals. Assuming the same cosmology and CAMB power spectra as before, we integrate from $q_i = 10^{-4}$ $h/$Mpc to $q_f=2$ $h/$Mpc at $z=0$. We form an equally spaced, 50x50 grid for $(k_{\bot}, k_z)$ with values between $0.001$ $h/$Mpc and $0.5$ $h/$Mpc including both ends. 

Fig.~ \ref{figure:power_theory_vs_sim} shows the theoretical $P_{\kappa}(k_{\bot}, k_z)$ on the bottom panel and a representative simulation on the top panel. While their structures and shapes agree, the theoretical $\kappa$ power spectrum is off by one order of magnitude. As we have discussed in the previous section, our theoretical $P_{\kappa_{3D}}$ estimate lacks significant power from reconstructed field because of the lowest order expansion. Since $P_{\kappa} = P^2_{\kappa_{3D}\delta}/ P_{\kappa_{3D}}$, the lack of power in $\kappa_{3D}$ results in high theoretical $P_{\kappa}$.

After confirming our theoretical framework in anisotropic power spectrum, we assess which modes contribute the most to the reconstructed field $\kappa$. We start by rewriting the cross spectrum.
\begin{equation}
	P_{\kappa_{3D}\delta} = \int_{\ln q_i}^{\ln q_f} q^{\prime} d\ln q^{\prime} \int_{-1}^1 dv^{\prime} \; I(q^{\prime}, v^{\prime})
\end{equation}
Because $I(q, v)$ is not symmetric with respect to $v$ (i.e. $I(q, v) \neq I(q, -v)$), we define $\bar I(q, v) = I(q, v) + I(q, -v)$ such that $v \in [0, 1]$. To quantify the importance of modes, we integrate the cross spectrum from $(q_i,0^\circ)$ to $(q, \theta)$, then divide this integral by the true value.
\begin{equation}
	\alpha(q, \theta) = \frac{1}{P_{\kappa_{3D}\delta}(k_{\bot}, k_{z})}  \int_{\ln q_{i}}^{\ln q} q^{\prime} d\ln q^{\prime} \int_{v}^{1} dv^{\prime} \; \bar I(q^{\prime}, v^{\prime})
\end{equation}

We evaluate the ratio $\alpha(q, \theta)$ at a fixed $(k_{\bot} = 0.05 h/$Mpc$, k_z = 0.02 h/$Mpc$)$ value at $z=0$ on a grid equally sampled with 100 $q$ points in $(10^{-4}, 1.5]$ $h/$Mpc and 90 equally spaced $\theta$ points in $(0, 90]^\circ$. $P_{\kappa_{3D}\delta}$ is evaluated with $q_f=2.5$ $h/$Mpc.

\begin{figure}
	\includegraphics[width=\columnwidth]{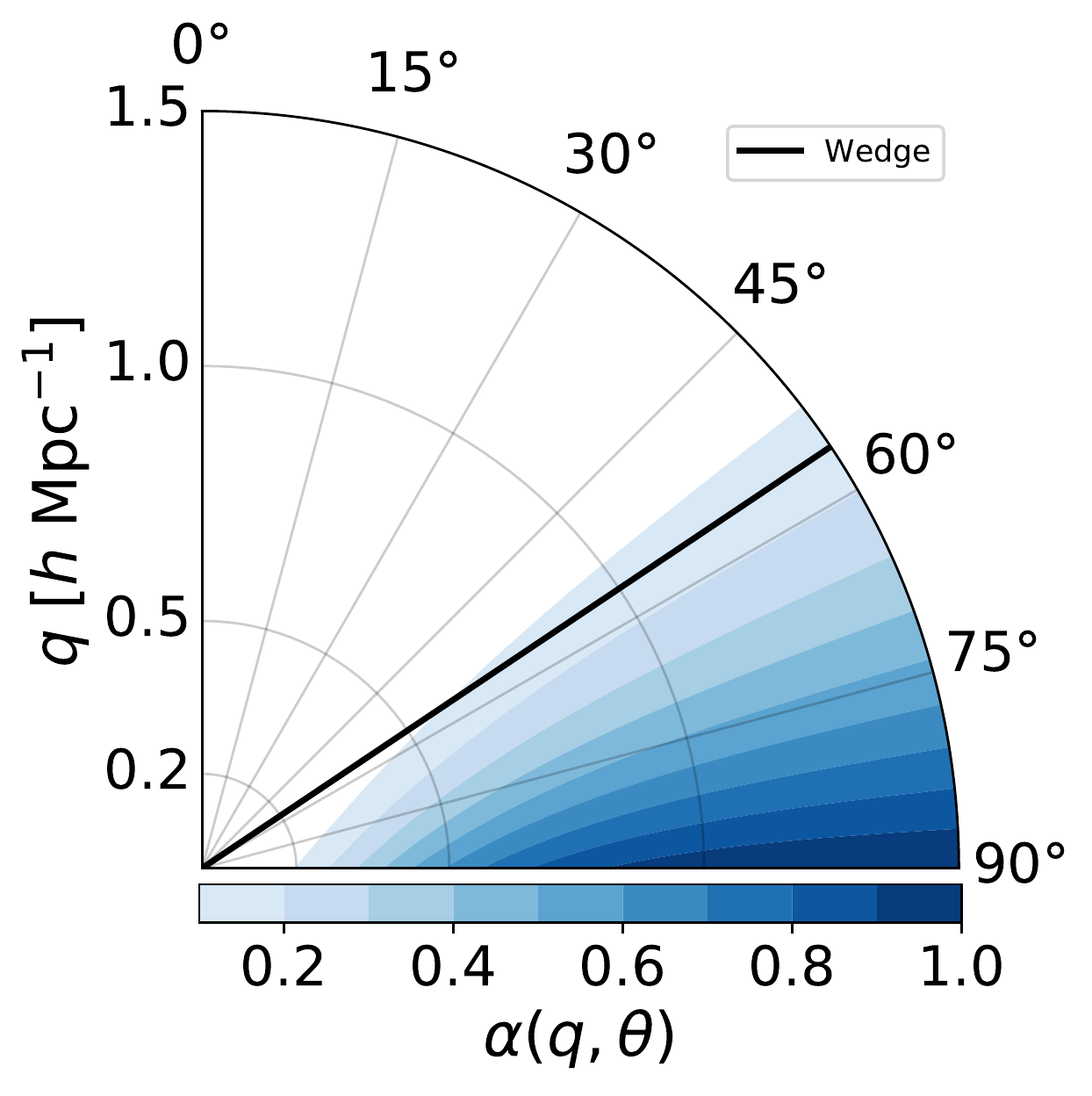}
	\caption{The fraction of cross correlation signal $\alpha(q, \theta)$ that can be achieved using modes up to $q$ and angle $\theta$ with respect to the line-of-sight at $z=0$. The reconstruction prefers $q_{\bot}$ modes and reaches its peak efficiency approximately at $(1.2$ $h/$Mpc$, \pi/2)$ Foreground wedge would restrict the reconstruction efficiency to approximately $0.2$.}
	\label{figure:theory_polar}
\end{figure}

The value of a point $(q,\theta)$ in Fig.~\ref{figure:theory_polar} represents the fraction of cross correlation signal that can be achieved using modes up to $q$ and angle $\theta$ with respect to the line-of-sight. The reconstruction prefers $q_{\bot}$ modes and reaches its peak efficiency at $q\approx 1.2$ $h/$Mpc when modes up to $\theta = \pi/2$ are accessible as the smallest fluctuations are suppressed by Gaussian smoothing. We added the wedge line for $z=1$ for reference. Foreground wedge impedes access to higher $\theta$ modes and restricts the efficiency of reconstruction to approximately $0.2$. Redshift space distortions affect the reconstruction in reverse direction, contaminating low $\theta$ modes. These are not significant in the reconstruction, explaining the weak dependence in Fig.~\ref{figure:compare_real_rsd_nofg_compare_kperp_real_z1}.

\section{Summary}
\label{sec:summary}
21-cm intensity mapping surveys lose large-scale radial signal to astrophysical smooth foregrounds. These foregrounds then contaminate smaller-scale radial and angular modes. Recovering the lost data (low $k_{\|}$ modes) will improve BAO measurements and cross correlations of 21-cm signal with CMB measurements and photo-z galaxy surveys. 

The large-scale over-density field deforms the local universe around an observer through a non-zero second derivative of the long-wavelength gravitational potential. This deformation imprints anisotropic features in the local small-scale power spectrum and allows us to reconstruct the lost data.

The second derivative of $\Phi_L$ gives the tidal field $t_{ij}(\tau) \equiv \Phi_{L, ij}(\textbf{0}) - \delta^K_{ij}\Phi_{L, kk}(\textbf{0})/3$. We choose two components ($\gamma_1$ and $\gamma_2$) to minimize redshift space distortions on the reconstruction. The estimators for each $\gamma$ can be constructed assuming the observed $\delta$ is a Gaussian field. In return, this assumption requires us to Gaussianize the observed $\delta$. The three dimensional convergence field $\kappa_{3D}$ is a linear combination of $\gamma_1$ and $\gamma_2$; and it is a better estimate for the underlying large-scale over-density field. We correct $\kappa_{3D}$ for bias and noise with a Wiener filter constructed from ten N-body simulations.

In this work, we reviewed non-linear tidal interactions using Lagrangian perturbation theory and explored the efficiency of reconstructing long-wavelength modes from local small-scale fluctuations. We performed new tests on cosmic tidal reconstruction by adding redshift space distortions, investigating a range of $k_\|$ and $k_{\bot}$ data loss and foreground wedge data loss. We also presented a novel theoretical framework to study the reconstruction, which can predict its efficiency and produce theoretical estimates for Wiener filter if improved.

We found the cross-correlation coefficient $r$ between the reconstructed field and true over-density field is above $0.7$ until $k=0.1$ $h/$Mpc in both spaces at two redshifts. Our numerical tests also confirmed the reconstruction is robust against peculiar velocities with minor degradation (less than 5\%) in $r$. This validates choosing quadrupolar distortions in the plane perpendicular to the line-of-sight, $\gamma_1$ and $\gamma_2$.

To assess 21-cm intensity mapping challenges, we tested the reconstruction against astrophysical foregrounds. We modelled spectrally smooth foregrounds as a sharp cutoff below $k^c_{\|}$ and found that the cross-correlation coefficient decreases by 20\%, but it is not sensitive to the cutoff value when $k^c_{\|} < 0.1$ $h/$Mpc. However, tidal reconstruction does not recover modes isotropically and specifically does not recover low $k_{\bot}$-high $k_{\|}$ modes. Given this feature, the reconstruction recovers modes $k_{\|} < 0.1$ $h/$Mpc up to $k_{\bot} = 0.2$ $h/$Mpc. Moreover, smooth foregrounds will leak into higher $k_{\bot}-k_{\|}$ modes due to imperfect instrumentation. We modelled this foreground wedge by removing modes obeying $|k_{\|}| \leq k_{\bot} D_M(z)H(z)/(1+z)$. After these modes are subtracted, $k^c_{\|}$ becomes irrelevant and the reconstruction deteriorates to $r(k<0.1 h/$Mpc$) \la 0.5$. In the anisotropic picture, modes around the wedge slope can be recovered with $r \sim 0.5$. On $k_\|=0$ plane, the wedge case performance shows mild improvement, whereas other cases do significantly better.

Finally, we incorporated all the theoretical segments into one framework, which showed that the cross correlations arise from a modified bispectrum and the power spectrum of $\kappa_{3D}$ originates from the four-point function. Our framework revealed a similar structure in $P_{\kappa}$, but missed the reconstructed signal due to lowest order expansion in the four-point function. Using the same framework, we estimated what modes are most important for efficient reconstruction. Our theoretical investigation predicted performance variations: steep decline when wedge modes are lost and minor degradation in redshift space. We expect one-loop corrections will decrease the discrepancy between simulations and theoretical predictions; it could construct a capable Wiener filter as well.

We have investigated the angular resolution and foregrounds individually. The available modes will depend on the details of the instrument. An exact treatment should be calculated into $P_N$ for real data applications. Since the modes above the wedge are less valuable, this should still produce correlations $r \sim 0.4$. 

Even though combining low angular resolution and the foreground wedge will not erase the correlations completely, we find the overall performance not as high as we would like. These challenges hint at utilizing the distortions in the radial direction. The trade-off for using all $t_{ij}$ components is to extend the study to redshift space. Since small scales matter, the instrument design should take into account of resolving smaller scales and mitigating foreground wedge as much as possible as well. We hope that our work stresses the necessity and importance of such further investigations. 

Throughout our work we assumed neutral hydrogen perfectly traces dark matter. The neutral hydrogen bias should be modelled for further investigation. A simple proposal would be to take haloes as tracers. 

\section*{Acknowledgements}
N. G. K. and N. P. are partially supported by the DOE DE-SC0017660.

We thank Dongzi Li, Hongming Zhu and Ue-Li Pen for helpful discussion.


\bibliographystyle{mnras}
\bibliography{Tidal_Refs}

\begin{thebibliography}{}
\makeatletter
\relax
\def\mn@urlcharsother{\let\do\@makeother \do\$\do\&\do\#\do\^\do\_\do\%\do\~}
\def\mn@doi{\begingroup\mn@urlcharsother \@ifnextchar [ {\mn@doi@}
  {\mn@doi@[]}}
\def\mn@doi@[#1]#2{\def\@tempa{#1}\ifx\@tempa\@empty \href
  {http://dx.doi.org/#2} {doi:#2}\else \href {http://dx.doi.org/#2} {#1}\fi
  \endgroup}
\def\mn@eprint#1#2{\mn@eprint@#1:#2::\@nil}
\def\mn@eprint@arXiv#1{\href {http://arxiv.org/abs/#1} {{\tt arXiv:#1}}}
\def\mn@eprint@dblp#1{\href {http://dblp.uni-trier.de/rec/bibtex/#1.xml}
  {dblp:#1}}
\def\mn@eprint@#1:#2:#3:#4\@nil{\def\@tempa {#1}\def\@tempb {#2}\def\@tempc
  {#3}\ifx \@tempc \@empty \let \@tempc \@tempb \let \@tempb \@tempa \fi \ifx
  \@tempb \@empty \def\@tempb {arXiv}\fi \@ifundefined
  {mn@eprint@\@tempb}{\@tempb:\@tempc}{\expandafter \expandafter \csname
  mn@eprint@\@tempb\endcsname \expandafter{\@tempc}}}

\bibitem[\protect\citeauthoryear{Adshead \& Furlanetto}{Adshead \&
  Furlanetto}{2008}]{adshead_reionization_2008}
Adshead P.~J.,  Furlanetto S.~R.,  2008, \mn@doi [Monthly Notices of the Royal
  Astronomical Society] {10.1111/j.1365-2966.2007.12681.x}, 384, 291

\bibitem[\protect\citeauthoryear{Akitsu \& Takada}{Akitsu \&
  Takada}{2018}]{akitsu_impact_2018}
Akitsu K.,  Takada M.,  2018, \mn@doi [Physical Review D]
  {10.1103/PhysRevD.97.063527}, 97, 063527

\bibitem[\protect\citeauthoryear{Akitsu, Takada  \& Li}{Akitsu
  et~al.}{2017}]{akitsu_large-scale_2017}
Akitsu K.,  Takada M.,   Li Y.,  2017, \mn@doi [Physical Review D]
  {10.1103/PhysRevD.95.083522}, 95, 083522

\bibitem[\protect\citeauthoryear{Bandura et~al.,}{Bandura
  et~al.}{2014}]{bandura_canadian_2014}
Bandura K.,  et~al., 2014. International Society for Optics and Photonics, p.
  914522, \mn@doi{10.1117/12.2054950}

\bibitem[\protect\citeauthoryear{Bull, Ferreira, Patel  \& Santos}{Bull
  et~al.}{2015}]{bull_late-time_2015}
Bull P.,  Ferreira P.~G.,  Patel P.,   Santos M.~G.,  2015, \mn@doi [The
  Astrophysical Journal] {10.1088/0004-637X/803/1/21}, 803, 21

\bibitem[\protect\citeauthoryear{Furlanetto \& Lidz}{Furlanetto \&
  Lidz}{2007}]{furlanetto_cross-correlation_2007}
Furlanetto S.~R.,  Lidz A.,  2007, \mn@doi [The Astrophysical Journal]
  {10.1086/513009}, 660, 1030

\bibitem[\protect\citeauthoryear{Jain \& Bertschinger}{Jain \&
  Bertschinger}{1994}]{jain_second-order_1994}
Jain B.,  Bertschinger E.,  1994, \mn@doi [The Astrophysical Journal]
  {10.1086/174502}, 431, 495

\bibitem[\protect\citeauthoryear{Jenkins}{Jenkins}{2010}]{jenkins_second-order_2010}
Jenkins A.,  2010, \mn@doi [Monthly Notices of the Royal Astronomical Society]
  {10.1111/j.1365-2966.2010.16259.x}, 403, 1859

\bibitem[\protect\citeauthoryear{Kaiser}{Kaiser}{1992}]{kaiser_weak_1992}
Kaiser N.,  1992, \mn@doi [The Astrophysical Journal] {10.1086/171151}, 388,
  272

\bibitem[\protect\citeauthoryear{Lu \& Pen}{Lu \&
  Pen}{2008}]{lu_precision_2008}
Lu T.,  Pen U.-L.,  2008, \mn@doi [Monthly Notices of the Royal Astronomical
  Society] {10.1111/j.1365-2966.2008.13524.x}, 388, 1819

\bibitem[\protect\citeauthoryear{Masui et~al.,}{Masui
  et~al.}{2013}]{masui_measurement_2013}
Masui K.~W.,  et~al., 2013, \mn@doi [The Astrophysical Journal]
  {10.1088/2041-8205/763/1/L20}, 763, L20

\bibitem[\protect\citeauthoryear{McCullagh, Neyrinck, Norberg  \&
  Cole}{McCullagh et~al.}{2016}]{mccullagh_recovering_2016}
McCullagh N.,  Neyrinck M.,  Norberg P.,   Cole S.,  2016, \mn@doi [Monthly
  Notices of the Royal Astronomical Society] {10.1093/mnras/stw223}, 457, 3652

\bibitem[\protect\citeauthoryear{Morales, Hazelton, Sullivan  \&
  Beardsley}{Morales et~al.}{2012}]{morales_four_2012}
Morales M.~F.,  Hazelton B.,  Sullivan I.,   Beardsley A.,  2012, \mn@doi [The
  Astrophysical Journal] {10.1088/0004-637X/752/2/137}, 752, 137

\bibitem[\protect\citeauthoryear{Newburgh et~al.,}{Newburgh
  et~al.}{2016}]{newburgh_hirax:_2016}
Newburgh L.~B.,  et~al., 2016. International Society for Optics and Photonics,
  p. 99065X, \mn@doi{10.1117/12.2234286}

\bibitem[\protect\citeauthoryear{Neyrinck, Szapudi  \& Szalay}{Neyrinck
  et~al.}{2011}]{neyrinck_rejuvenating_2011}
Neyrinck M.~C.,  Szapudi I.,   Szalay A.~S.,  2011, \mn@doi [The Astrophysical
  Journal] {10.1088/0004-637X/731/2/116}, 731, 116

\bibitem[\protect\citeauthoryear{Pen, Sheth, Harnois-Deraps, Chen  \& Li}{Pen
  et~al.}{2012}]{pen_cosmic_2012}
Pen U.-L.,  Sheth R.,  Harnois-Deraps J.,  Chen X.,   Li Z.,  2012, arXiv
  preprint arXiv:1202.5804

\bibitem[\protect\citeauthoryear{Schmidt, Pajer  \& Zaldarriaga}{Schmidt
  et~al.}{2014}]{schmidt_large-scale_2014}
Schmidt F.,  Pajer E.,   Zaldarriaga M.,  2014, \mn@doi [Physical Review D]
  {10.1103/PhysRevD.89.083507}, 89, 083507

\bibitem[\protect\citeauthoryear{Scoccimarro}{Scoccimarro}{1998}]{scoccimarro_transients_1998}
Scoccimarro R.,  1998, \mn@doi [Monthly Notices of the Royal Astronomical
  Society] {10.1046/j.1365-8711.1998.01845.x}, 299, 1097

\bibitem[\protect\citeauthoryear{Seo \& Hirata}{Seo \&
  Hirata}{2016}]{seo_foreground_2016}
Seo H.-J.,  Hirata C.~M.,  2016, \mn@doi [Monthly Notices of the Royal
  Astronomical Society] {10.1093/mnras/stv2806}, 456, 3142

\bibitem[\protect\citeauthoryear{Springel}{Springel}{2005}]{springel_cosmological_2005}
Springel V.,  2005, \mn@doi [Monthly Notices of the Royal Astronomical Society]
  {10.1111/j.1365-2966.2005.09655.x}, 364, 1105

\bibitem[\protect\citeauthoryear{Springel, Yoshida  \& White}{Springel
  et~al.}{2001}]{springel_gadget:_2001}
Springel V.,  Yoshida N.,   White S. D.~M.,  2001, \mn@doi [New Astronomy]
  {10.1016/S1384-1076(01)00042-2}, 6, 79

\bibitem[\protect\citeauthoryear{Weinberg}{Weinberg}{1992}]{weinberg_reconstructing_1992}
Weinberg D.~H.,  1992, \mn@doi [Monthly Notices of the Royal Astronomical
  Society] {10.1093/mnras/254.2.315}, 254, 315

\bibitem[\protect\citeauthoryear{White \& Padmanabhan}{White \&
  Padmanabhan}{2017}]{white_matched_2017}
White M.,  Padmanabhan N.,  2017, \mn@doi [Monthly Notices of the Royal
  Astronomical Society] {10.1093/mnras/stx1682}, 471, 1167

\bibitem[\protect\citeauthoryear{Zhu, Pen, Yu, Er  \& Chen}{Zhu
  et~al.}{2016}]{zhu_cosmic_2016}
Zhu H.-M.,  Pen U.-L.,  Yu Y.,  Er X.,   Chen X.,  2016, \mn@doi [Physical
  Review D] {10.1103/PhysRevD.93.103504}, 93, 103504

\bibitem[\protect\citeauthoryear{Zhu, Pen, Yu  \& Chen}{Zhu
  et~al.}{2018}]{zhu_recovering_2018}
Zhu H.-M.,  Pen U.-L.,  Yu Y.,   Chen X.,  2018, \mn@doi [Physical Review D]
  {10.1103/PhysRevD.98.043511}, 98, 043511

\makeatother
\end{thebibliography}



\appendix

\section{Time Dependent Functions}
\label{app:time_dep_fn}
The redshift dependent function $f$ is given by 
\begin{equation}
	f(k, z) = -2D_{1st}(z) + F(z)\left(2 - \frac{d \ln P_{l}(k, z)}{d\ln k}\right),
\end{equation}
where $D(z)$ is the linear growth function and
\begin{align}
	F(z) &= \int_{z}^{\infty} \frac{D(z^{\prime \prime}) d z^{\prime \prime}}{H(z^{\prime \prime})}  \int_{z}^{z^{\prime \prime}} \frac{(1 + z^{\prime})d z^{\prime}}{H(z^{\prime})} \\
    D_{1st}(z) &= \int_{z}^{\infty} \frac{Y(z^\prime) d z^{\prime}}{H(z^{\prime})^{2}}\left[\frac{H(z)}{D(z)} D(z^{\prime}) - H(z^{\prime}) \right].
\end{align}
We defined the following intermediate functions to simplify expressions,
\begin{align}
	T(z) &= D(z) (1 + z) \\
    W(z) &= H(z) D^{\prime}(z) - H^{\prime}(z) D(z)\\
    Y(z) &= T(z) D(z) / W(z).
\end{align}

\section{Details on Reconstruction Theory}
\label{app:reconst_theory}
Three dimensional convergence field is given by the integration in equation~(\ref{eq:analytic_kappa3d}). We start by writing $\tilde{T}_{\kappa_{3D}}$.
\begin{align}
	\tilde{T}_{\kappa_{3D}}(\textbf{k}^{\prime}, \textbf{k} - \textbf{k}^{\prime}) &= \frac{2k^{2}}{3(k_{x}^{2} + k_{y}^{2})^{2}} \\
	&\times \left[  (k_{x}^{2}-k_{y}^{2}) \tilde{T}_{\gamma_{1}}(\textbf{k}^{\prime}, \textbf{k} - \textbf{k}^{\prime}) + 2k_{x}k_{y} \tilde{T}_{\gamma_{2}}(\textbf{k}^{\prime}, \textbf{k} - \textbf{k}^{\prime})\right]\nonumber
\end{align}
Since $T_{\gamma_{i}}$ is symmetric, $\tilde{T}_{\kappa_{3D}}(\textbf{k}^{\prime}, \textbf{k} - \textbf{k}^{\prime})$ is symmetric as well.
\begin{equation}
	\tilde{T}_{\kappa_{3D}}(\textbf{k}^{\prime}, \textbf{k} - \textbf{k}^{\prime}) = \tilde{T}_{\kappa_{3D}}(\textbf{k} - \textbf{k}^{\prime}, \textbf{k}^{\prime})
\end{equation}
We can reshape this expression into the following:
\begin{align}
	\tilde{T}_{\kappa_{3D}}(\textbf{q}, \textbf{k} - \textbf{q})&= -\frac{2k^{2}w(\textbf{q}) w(\textbf{k} - \textbf{q})}{3k^{4}_{\bot}} \\
	&\times \left[  \frac{k^{2}_{\bot}q^{2}_{\bot} - 2(\textbf{k}_{\bot}\cdot\textbf{q}_{\bot})^{2} + (\textbf{k}_{\bot}\cdot\textbf{q}_{\bot})k^{2}_{\bot}}{q |\textbf{k}-\textbf{q}|} \right].\nonumber
\end{align}

We want to evaluate $\tilde{T}_{\kappa_{3D}}(\textbf{q}, \textbf{k} - \textbf{q})$ in spherical coordinates. We start by fixing $\textbf{k}_{\bot}$ and assigning an angle $\phi$ between $\textbf{k}_{\bot}$ and $\textbf{q}_{\bot}$. We continue using $\textbf{p} \equiv \textbf{k} - \textbf{q}$ and define $t\equiv\cos\phi$. Then, the following relations hold:
\begin{align}
	\textbf{k}\cdot \textbf{q} &= k_{\bot}q_{\bot}t + k_{z} q_{z}\\
	p_{\bot} &= \sqrt{k^{2}_{\bot} + q_{\bot}^{2} - 2k_{\bot}q_{\bot}t }\\
	p_{z} &= k_{z} - q_{z} \\
	p &= \sqrt{k^{2}+q^{2}- 2\textbf{k}\cdot \textbf{q} }.
\end{align}
We further define $q_{z} \equiv q \cos\theta = q v$. Then $q_{\bot} = q \sqrt{1-v^{2}}$. We can rewrite the relations above in terms of these new spherical coordinates.
\begin{align}
	\textbf{k}\cdot \textbf{q} &= k_{\bot}q t\sqrt{1-v^{2}} + k_{z} q v\\
	p_{\bot} &= \sqrt{k^{2}_{\bot} + q^{2}(1-v^{2}) - 2k_{\bot}q t \sqrt{1-v^{2}}}\\
	p_{z} &= k_{z} - q v
\end{align}

Finally, $\tilde{T}_{\kappa_{3D}}$ can be expressed in terms of the same coordinates.
\begin{align}
	\tilde{T}_{\kappa_{3D}}(q, v, t, k_{\bot}, k_{z}) &= -\frac{2k^{2}}{3k^{2}_{\bot}}  \frac{ w(q_{\bot}, q_{z})w(p_{\bot}, p_{z})}{p} \\
	&\times [q(1-v^{2}) (1-2t^{2}) + k_{\bot} t\sqrt{1-v^{2}}]\nonumber
\end{align}

\section{Smoothing}
\label{app:smooth}
Let us consider the smoothing: $\delta_{R}(\textbf{k}) = S_{R}(k) \delta(\textbf{k})$. Then,
 \begin{align}
	 \delta_{G}(\textbf{k})&= \delta_{R}(\textbf{k}) - \frac{1}{2} \int \frac{d^{3}q}{(2\pi)^{3}} \delta_{R}(\textbf{q}) \delta_{R}(\textbf{k} - \textbf{q}) + \Theta(\delta^{3})\\
	 &= S_{R}(k) \delta(\textbf{k})- \frac{1}{2} \int \frac{d^{3}q}{(2\pi)^{3}} S_{R}(q) S_{R}(p)\delta(\textbf{q}) \delta(\textbf{p})\\
	 &=  S_{R}(k)\left( \delta_{l}(\textbf{k}) +\int \frac{d^{3}q}{(2\pi)^{3}}  F_{2}(\textbf{q}, \textbf{p}) \delta_{l}(\textbf{q}) \delta_{l}(\textbf{p}) \right) \\
	 &\quad- \frac{1}{2} \int \frac{d^{3}q}{(2\pi)^{3}} S_{R}(q) S_{R}(p)\delta_l(\textbf{q}) \delta_l(\textbf{p})\nonumber\\
	 \delta_{G}(\textbf{k}) &= S_{R}(k) \left(\delta_{l}(\textbf{k}) +  \int \frac{d^{3}q}{(2\pi)^{3}}  F_{2, G, R}(\textbf{q}, \textbf{p}) \delta_{l}(\textbf{q}) \delta_{l}(\textbf{p})\right),
\end{align}
where we have modified the kernel to include smoothing.
\begin{equation}
	 F_{2, G, R}(\textbf{k}_{1}, \textbf{k}_{2}) = F_{2}(\textbf{k}_{1}, \textbf{k}_{2})- \frac{1}{2} \frac{S_{R}(k_{1}) S_{R}(k_{2})}{S_{R}(|\textbf{k}_{1} + \textbf{k}_{2}|)}
\end{equation}

There are two modifications to previous modified bispectrum $B_T$ expression. First, we multiply $B_T$ by $S_{R}(q) S_{R}(p)$. Second, we substitute $F_{2, G} \rightarrow F_{2, G, R}$.

\begin{align}
	B_{T, R}&= S_{R}(q) S_{R}(p)2 [ F_{2}(\textbf{q}, \textbf{p}) P_{l}(q) P_{l}(p) \\
	&\qquad\qquad\qquad + F_{2, G, R}(\textbf{p}, -\textbf{k}) P_{l}(p) P_{l}(k) \nonumber \\
	&\qquad\qquad\qquad + F_{2, G, R}(\textbf{q}, - \textbf{k}) P_{l}(q) P_{l}(k) ] \nonumber\\
	B_{T, R}(\textbf{q}, \textbf{p}) &= S_{R}(q) S_{R}(p) \\
	&\times \left[B -  \frac{S_{R}(p) S_{R}(k)}{S_{R}(q)}  P_{l}(p) P_{l}(k) - \frac{S_{R}(q) S_{R}(k)}{S_{R}(p)}  P_{l}(q) P_{l}(k) \right]\nonumber
\end{align}


\bsp	
\label{lastpage}
\end{document}